\documentclass[journal]{IEEEtran}
\usepackage{tikz}
\usepackage{float}
\usepackage{mcite}
\usepackage{graphicx}
\usepackage{amsmath}
\usepackage{amssymb}
\usepackage{bm}
\usepackage{upgreek}
\usepackage{mathrsfs}
\usepackage[colorlinks=true, allcolors=blue]{hyperref}
\usepackage[capitalize]{cleveref}
\usepackage{subfigure}
\usepackage{array}
\usepackage{threeparttable}
\usepackage{booktabs}
\usepackage{bbding}
\usepackage{algorithm}  
\usepackage{algpseudocode}

\definecolor{linkcolor}{RGB}{92,92,192}
\definecolor{blackcolor}{RGB}{0,0,0}
\hypersetup{colorlinks=true,linkcolor=blackcolor,citecolor=blackcolor,urlcolor=linkcolor}

\IEEEoverridecommandlockouts

\begin{document}

%\title{Residual-to-Residual Deep Neural Network Series for Non-Cartesian Magnetic Resonance Imaging}

%\title{R2D2 Deep Neural Network Series for Scalable Non-Cartesian Magnetic Resonance Imaging}

%\title{R2D2 DNN Series for Scalable Non-Cartesian MRI}

\title{Interlaced R2D2 DNN Series for Scalable Non-Cartesian MRI with Sensitivity Self-calibration}

\author{\IEEEauthorblockN{
%Yiwei Chen$^1$, Amir Aghabiglou$^1$, Motahare Torki$^1$, Shijie Chen$^1$, Chao Tang$^{1,3}$, %Chung San Chu$^1$, Ruud B. van Heeswijk$^2$,
S. Chen$^1$, Y. Chen$^1$, A. Aghabiglou$^1$, M. Torki$^1$, C. Tang$^{1}$,  R. B. van Heeswijk$^2$, 
Y. Wiaux$^1$ \IEEEauthorrefmark{1}\thanks{%The work was supported by EPSRC under grants EP/T028270/1 and ST/W000970/1. Computing \mbox{resources} came from the Cirrus UK National Tier-2 HPC Service at EPCC (http://www.cirrus.ac.uk) funded by the University of Edinburgh and EPSRC (EP/P020267/1). %, partly through time allocation under the SUSA project, and partly through GPU resources directly provided by EPCC. A.~J.~was supported by EPSRC under grant EP/T028351/1.
}}
\IEEEauthorblockA{\\$^1$Institute of Sensors, Signals and Systems, Heriot-Watt University Edinburgh, United Kingdom 
% \\$^2$EPCC, University of Edinburgh, %Edinburgh EH8 9BT, United Kingdom
\\$^2$Department of Diagnostic Imaging and Interventional Radiology, Lausanne University and Hospital, Switzerland
}
\\
\IEEEauthorblockA{Email: \IEEEauthorrefmark{1}y.wiaux@hw.ac.uk}
}

\maketitle

\begin{abstract}
We introduce interlaced R2D2 (iR2D2), a DNN series paradigm for scalable image reconstruction from accelerated non-Cartesian k-space acquisitions in MRI with sensitivity map self-calibration. While unrolled DNN architectures provide robust image formation, embedding non-uniform fast Fourier transform operators within the backpropagation graph becomes impractical to train at large scale, \emph{e.g.},~in 2D MRI with a large number of coils, or for higher-dimensional imaging. 
To address this scalability challenge, we leverage the R2D2 paradigm as a learned version of the Matching Pursuit algorithm that was recently introduced in radio astronomy for fast large-scale Fourier imaging. R2D2’s reconstruction is formed as a series of residual images iteratively estimated as outputs of DNN modules taking the previous iteration’s data residual as input. 
Specific to MRI, precomputed sensitivity maps derived from undersampled data can yield an inaccurate measurement operator, which may adversely affect the performance of iterative algorithms such as R2D2.
Thus, we extend the R2D2 framework to iR2D2 by introducing
a bespoke interlaced architecture that alternates between two R2D2 DNN series to jointly self-calibrate sensitivity maps and form the MR image. 
We further enhance iR2D2 to operate as an adaptive solver governed by an error-controlled update condition that enforces a sufficient residual energy descent, a dynamic capability fundamentally incompatible with the predefined forward passes of unrolled architectures. 
Extensive experiments in simulation and on real data, targeting undersampled radial k-space sampling, demonstrate that iR2D2 significantly improves upon R2D2 and outperforms state-of-the-art benchmarks, delivering scalable, high-fidelity imaging with corrected sensitivity profiles.

\end{abstract}


\IEEEpeerreviewmaketitle

\section{Introduction}
Computational imaging holds substantial scientific attention due to its broad spectrum of applications in disciplines such as astronomy \cite{wiaux2009compressed} and medicine \cite{glover2011overview}. Among these, Magnetic Resonance Imaging (MRI) stands out as a representative technique, enabling high-precision reconstruction of anatomical structures and organs, thereby serving as an indispensable tool in clinical diagnostics \cite{liang2000principles}. MRI acquires signals in the spatial frequency domain (k-space) by leveraging the interaction between magnetic fields and radio waves. Since fully sampling k-space requires long scanning times, measurements are typically undersampled \cite{qu2012undersampled}, rendering the inverse problem ill-posed. To mitigate this, Parallel Imaging (PI) exploits spatial sensitivity variations across multiple receiver coils \cite{pruessmann1999sense}, while Compressed Sensing (CS) leverages sparsity priors via handcrafted regularization \cite{lustig2007sparse}. Building on this, classical optimization approaches recover the image by balancing data consistency with a prior regularization term \cite{knoll2020deep}. These priors--such as the $\ell_1$-norm of the image decomposed in a wavelet basis \cite{guerquin2011fast} or its gradient magnitude \cite{block2007undersampled}--must be carefully handcrafted and tailored to each specific inverse problem, requiring significant domain expertise. 
Furthermore, reconstructing high-fidelity images from accelerated data using these iterative approaches is computationally demanding and scales poorly with the large data volumes inherent to high-density coil arrays or high-dimensional imaging \cite{stankovic20144d}. 

Deep Neural Networks (DNNs) have fundamentally shifted this paradigm. Pure DNNs \cite{ronneberger2015u, hyun2018deep, han2019k, eo2018kiki} offer rapid inference but lack explicit physical constraints, failing to enforce data consistency. To address this, unrolled DNNs (such as PD-Net \cite{adler2018learned}) mimic optimization iterations within their architecture to explicitly enforce data consistency. Similarly, Plug-and-Play (PnP) approaches \cite{ahmad2020plug} integrate a learned, data-driven denoiser within an iterative optimization algorithm to replace handcrafted regularization.
 
While effective for Cartesian data, these methods face severe scalability barriers when applied to undersampled non-Cartesian acquisitions. Non-Cartesian reconstruction relies on large-scale measurement operators based on the Non-Uniform Fast Fourier Transform (NUFFT). Unrolled architectures remain fundamentally tied to embedding this measurement operator directly within the training graph.
Even with recent optimizations--such as the 3D multi-coil NC-PDNet \cite{ramzi2022nc, NCPDNET2025}, which utilizes coil compression and limited unrolled iterations--the NUFFT imposes massive memory and computational demands during training, strictly limiting flexibility regarding operator complexity or dimensionality. 
Decomposed Diffusion Sampling (DDS) \cite{chung2024decomposed} eliminates the NUFFT from the training phase entirely by training a denoising DNN solely for the diffusion process. 
Conceptually resembling PnP, DDS alternates between a data-fidelity update and a denoising step, leveraging a powerful generative prior for superior regularization \cite{zhu2023denoising}. However, like PnP, DDS still requires hundreds of NUFFT-dependent inference steps, critically hindering scalability during reconstruction.

Recently, the ``\textbf{R}esidual-to-\textbf{R}esidual \textbf{D}NN series for high-\textbf{D}ynamic range imaging" (R2D2) approach, a learned variant of the Matching Pursuit algorithm \cite{mallat1993matching}, has demonstrated a breakthrough in balancing reconstruction quality and computational efficiency in radio-interferometric imaging \cite{aghabiglou2023deep, aghabiglou2024r2d2} and a proof of concept in single-coil, real-valued non-Cartesian MRI \cite{chen2024scalable}. 
R2D2 performs reconstruction as a sequence of residual image estimations. Each residual image is iteratively predicted by a DNN, which takes the previous image estimate and the associated back-projected data residual as input. 
Moreover, its number of iterations is significantly lower than the typical iteration count in PnP algorithms. R2D2 thus holds great potential for large-dimensional MRI, offering a promising direction for efficient and accurate imaging.  

Simultaneously, a critical limitation in current multi-coil MRI reconstruction is the reliance on precomputed sensitivity maps. Standard reconstruction assumes perfect knowledge of these maps, typically estimated via ESPIRiT. However, with undersampled data, sensitivity map calibration is often suboptimal, leading to forward model errors. Consequently, learning-based models are forced to rely on imperfect estimates during both training and inference, imposing a fundamental constraint on achievable reconstruction quality.

To bridge this gap between theoretical constraints and practical acquisition realities, we introduce iR2D2. Unlike classical approaches that precompute sensitivity maps prior to reconstruction \cite{uecker2014espirit}, deep learning architectures that update them using shared parameters \cite{ramzi2022nc}, or classical joint optimization techniques that rely on computationally intensive iterative solvers \cite{ying2007joint}, iR2D2 addresses the inverse problem through a learned, interlaced architecture. 
The framework seamlessly alternates between two parallel R2D2 DNN series--one that self-calibrates the sensitivity maps and another that performs image reconstruction--iteratively refining both the physical measurement operator and the anatomical estimate. 

Furthermore, capitalizing on the architectural flexibility of R2D2, we employ the U-WDSR network \cite{aghabiglou2025r2d2}, novel to the MRI domain, as the core image reconstruction module to achieve superior accuracy and robustness. Complementing this, we utilize a U-Net \cite{ronneberger2015u} architecture for the sensitivity estimation modules, ensuring robust capture of smooth coil profiles while maintaining computational efficiency. 

Moreover, we enhance the proposed iR2D2 architecture to operate as an adaptive optimization scheme \cite{bauschke2020correction}. While reminiscent of block-coordinate forward–backward (BCFB) algorithms utilized in classical joint calibration \cite{chouzenoux2016block, repetti2017non}--which allow multiple sequential updates per variable (sensitivity or image)--iR2D2 fundamentally generalizes this approach. Rather than relying on a fixed, pre-determined update schedule, iR2D2 dictates the block-coordinate sequence dynamically through an error-controlled  Update Condition (UC), grounded in a sufficient decrease of the residual energy. This algorithmic adaptability is also another significant advantage over the predefined forward passes of end-to-end unrolled networks.

We conducted an extensive evaluation on both simulated and real datasets using undersampled radial k-space acquisition. The results demonstrate that iR2D2 outperforms State-of-The-Art (SOTA) methods in both reconstruction quality and efficiency, offering a novel scalable paradigm for high-fidelity non-Cartesian MRI reconstruction.

%%%%%%%%%%%%%%%%%%%%%%%%%%%%%%%%%%%%%%%%%%
\section{Preliminaries}

\subsection{Non-Cartesian MRI}
Non-Cartesian MRI acquisitions offer advantages for advanced imaging applications \cite{wright2014non}. Contemporary MRI scanners utilize multiple receiver coils positioned around a region of interest to acquire multiple MR signals from that region. Each coil possesses a distinct sensitivity map that determines its effectiveness at various imaging locations. In the non-Cartesian context, the measurement process reads as
\begin{equation}
\bm{y}_{\ell} =\bm{\mathsf{\Phi}}_{\ell} \bm{x^{\star}}+ \bm{n}_{\ell}, \quad \forall \ell=1, \ldots, L,
\label{Eq:multicoil_MRI}
\end{equation}
where $\bm{x^{\star}} \in \mathbb{C}^N$ denotes the 2D complex-valued unknown MR image of interest, $\bm{y}_{\ell}$ and $\bm{n}_{\ell}$ are the k-space measurement and Gaussian random noise with mean zero and standard deviation $\tau_{\ell}$ of the $\ell$-th coil, respectively. $\bm{\mathsf{\Phi}}_{\ell} \triangleq \bm{\mathsf{F}}_{\Omega} \bm{\mathsf{s}}_{\ell}$ is the sampling operator based on the Non-Uniform Discrete Fourier Transform $\bm{\mathsf{F}}_{\Omega}$, with the k-space sampling location denoted by $\Omega$ and the diagonal matrix $\bm{\mathsf{s}}_{\ell} \in \mathbb{C}^{N \times N}$ representing the $\ell$-th coil sensitivity map. In practice, the approximation of $\bm{\mathsf{F}}_{\Omega}$ is achieved via the NUFFT \cite{fessler2003nonuniform}. %In the single-coil acquisition, $\bm{\mathsf{s}}_1$ is the identity matrix $\mathbf{I}$. 
Generally, the coil sensitivity maps are assumed to satisfy the constraint
\begin{equation}
	\sum_{\ell=1}^L \bm{\mathsf{s}}_{\ell}^{\dagger} \bm{\mathsf{s}}_{\ell}=
    \sum_{\ell=1}^L |\bm{\mathsf{s}}_{\ell}|^2=
    \mathbf{I},
    \label{Eq:sens}
\end{equation}
By enforcing this condition, any spatial deviations are inherently absorbed into the target Ground-Truth (GT) image $\bm{x^{\star}}$.

\subsection{Density compensation}
In contrast to the Fast Fourier Transform, the NUFFT is not invertible. Thus, the application of the adjoint of NUFFT to k-space data frequently results in outcomes that diverge considerably from the sought image. For instance, in radial sampling, the densely sampled regions proximal to the center of k-space tend to amass disproportionately large intensities, culminating in a back-projected image exhibiting unnaturally high values. To address this challenge, density compensation, as introduced by \cite{pipe1999sampling}, determines weighting factors in k-space that balance the contributions from various sample locations. This balance is achieved by iteratively applying both the interpolation matrix and its adjoint over iterations. After pre-calculating and storing the density compensation weights as the elements of the diagonal matrix $\bm{\mathsf{D}}$, multiplying the k-space data by $\bm{\mathsf{D}}$ before back-projection facilitates the generation of a more balanced and accurate back-projected image as
\begin{equation}
{\bm x}_{\rm b} = \kappa \sum_{\ell=1}^{L} \bm{\mathsf{s}}_{\ell}^{\dagger} \bm{\mathsf{F}}_{\Omega}^{\dagger} \bm{\mathsf{D}}\bm{y}_{\ell}=\kappa \sum_{\ell=1}^{L} \bm{\mathsf{\Phi}}_{\ell}^{\dagger}\bm{\mathsf{D}}\bm{y}_{\ell},
	\label{Eq:back_multi}
\end{equation}
where $(\cdot)^{\dagger}$ is the adjoint operator. $\kappa=\left[\max \left(\left|\sum_{\ell=1}^{L} \bm{\mathsf{\Phi}}^{\dagger}_{\ell}\bm{\mathsf{D}} \bm{\mathsf{\Phi}}_{\ell}\bm{\delta}\right|\right)\right]^{-1}$ is a normalization constant.
$\bm{\delta}$ represents a unit-magnitude Dirac impulse image with central element $\frac{\sqrt{2}}{2}(1 + i)$ and zeros elsewhere.

\subsection{Sensitivity estimation}
Coil sensitivity estimation is a critical prerequisite for achieving high-fidelity multi-coil MR image reconstruction. Unlike simple image-based approximations, such as normalizing individual coil images by a root-sum-of-squares reference, the ESPIRiT algorithm \cite{uecker2014espirit} derives smooth, spatially consistent sensitivity maps directly from the acquired k-space data. 
These estimated maps are then incorporated into the forward encoding operator to solve the inverse reconstruction problem. However, in highly accelerated non-Cartesian acquisitions, the calibration region is often limited or inconsistent, leading to imperfect sensitivity estimates. Such structural inaccuracies propagate through the reconstruction pipeline and ultimately constrain achievable image quality, making sensitivity estimation a key bottleneck under these conditions.

%%%%%%%%%%%%%%%%%%%%%%%%%%%%%%%%%%%%%%%%%%
\section{Algorithmic structure}
\label{sec:method}

\subsection{R2D2}
The R2D2 algorithm requires a series of $I$ DNN modules represented as $\{\bm{\mathsf{G}}_{\bm{\widehat{\theta}^{(i)}}}\}_{1\leq i \leq I}$, each with its own set of learnable parameters $\{\bm{\widehat{\theta}^{(i)}}\}_{1\leq i \leq I}$. Each DNN takes two input images: the previous image estimate $\bm{x}^{(i-1)}$ and its back-projected data residual
\begin{equation}
\bm{r}^{(i-1)} = \kappa \sum_{\ell=1}^{L} \bm{\mathsf{\Phi}}_{\ell}^{\dagger}\bm{\mathsf{D}}\left(\bm{y}_{\ell} - \bm{\mathsf{\Phi}}_{\ell} \bm{x}^{(i-1)}\right). 
\label{Eq:residual}
\end{equation}
The current image estimate is updated based on the output of the corresponding DNN module as 
\begin{equation}
    \bm{x}^{(i)} = \bm{x}^{(i-1)} + \bm{\mathsf{G}}_{\bm{\widehat{\theta}^{(i)}}}(\bm{r}^{(i-1)}, \bm{x}^{(i-1)}).
    \label{Eq:image_estimate}
\end{equation}
The image estimate is initialized as $\bm{x}^{(0)} = \bm{0}$, yielding an initial back-projected residual $\bm{r}^{(0)} = \bm{x}_{\rm b}$. This iterative structure allows for the progressive improvement of the resolution and structural details.  
The final reconstruction $\widehat{\bm{x}}$ is given by
\begin{equation}
    \widehat{\bm{x}} \triangleq \bm{x}^{(I)} = \sum_{i=1}^I \bm{\mathsf{G}}_{\bm{\widehat{\theta}^{(i)}}}(\bm{r}^{(i-1)}, \bm{x}^{(i-1)}). \label{Eq:final_image_estimate}
\end{equation}

R2D2 features a hybrid structure that combines aspects of learned matching pursuit. Unlike PnP algorithms and DDS, which utilize denoisers blind to the acquisition physics, the DNN modules in R2D2 are explicitly informed by the measurement model, learning to iteratively extract structural information directly from the back-projected data residuals. In contrast to unrolled DNN modules \cite{sriram2020end, ramzi2022nc}, R2D2 externalizes the back-projected data residual calculation from the network structure, which helps to alleviate the computational burden of NUFFT during training and improves scalability. Furthermore, although R2D2 is conceptually similar to hybrid strategies like greedy, stack-by-stack learning \cite{daval2023deep}--in decoupling data consistency from gradient computation--it avoids their key compromises. Prior methods rely on concatenated state buffers and heavily approximate the measurement operator via coil and model compression to maintain computational feasibility. In contrast, R2D2 retains the exact operator and propagates state solely through the physical residual \cref{Eq:residual} and previous image estimate \cref{Eq:image_estimate}, preserving theoretical clarity without introducing truncation errors.

\subsection{iR2D2}
Accurate coil sensitivity maps are critical for high-quality multi-coil MRI reconstruction, particularly in non-Cartesian acquisitions where calibration data are often imperfect. To bridge the gap between practical reconstruction with imperfect sensitivity estimates and ideal reconstruction with accurate sensitivity profiles, we propose iR2D2, an extension of R2D2 that sequentially interlaces the image reconstruction with coil sensitivity maps refinement during reconstruction. iR2D2 is also reminiscent of the block-coordinate forward–backward structure employed for joint calibration in \cite{repetti2017non}. iR2D2 employs a series of distinct DNN modules for this iterative refinement. Unlike NC-PDNet and R2D2-Net, which typically employ a single model for sensitivity map updating, iR2D2 leverages distinct network parameters for both image reconstruction and sensitivity map estimation at each iteration. This allows the framework to adapt its processing complexity to the evolving image features and residuals. 

Formally, in non-Cartesian multi-coil MRI, the forward model can be written as a bilinear inverse problem
\begin{equation}
\bm{y}_{\ell} = \bm{\mathsf{F}}_{\Omega}\left(\bm{\mathsf{s}}_{\ell} \odot \bm{x}^\star\right) + \bm{n}_{\ell}, \quad \forall \ell=1, \ldots, L.
\end{equation}
To solve this problem, iR2D2 adopts an alternating minimization strategy by interlacing two parallel series of DNN modules: one dedicated to sensitivity map self-calibration, and the other to image reconstruction.
\begin{figure*}[ht!]
\vspace{-5mm}
\includegraphics[width=\textwidth]{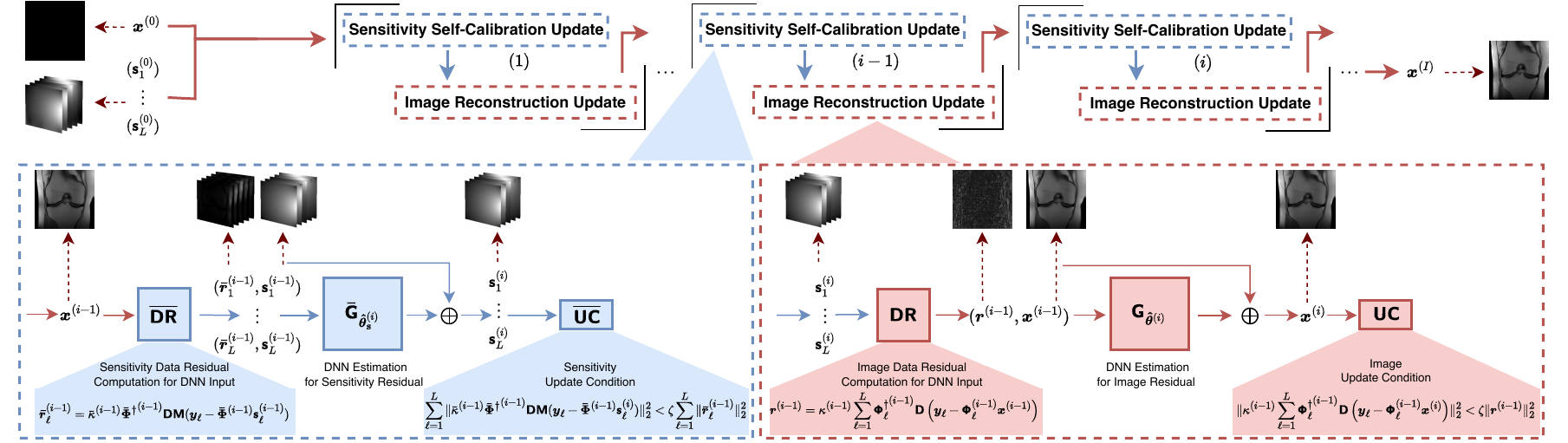}
\caption{Illustration of the iR2D2 algorithm. The diagram highlights the sequential evolution of sensitivity and image estimates. At each iteration, the sensitivity map estimation DNNs accept the previous estimate $\bm{\mathsf{s}}_{\ell}^{(i-1)}$ and the corresponding back-projected data residual $\bm{\bar{r}}_{\ell}^{(i-1)}$ to dynamically refine the coil profiles. Utilizing these corrected physical operators, the R2D2 image reconstruction DNNs accept the updated back-projected data residual $\bm{r}^{(i-1)}$ and the previous image estimate $\bm{x}^{(i-1)}$ to update the reconstructed image. The UC and $\overline{\text{UC}}$, detailed in Sec.~\ref{sec:convergence}, act as a conditional error-controlled criterion for both update paths. If the residual energy descent constraint is violated, the proposed update is rejected, and the previous state is propagated forward.
}
\vspace{-5mm}
\label{fig:jcr2d2}
\end{figure*}
When updating the image, fixing the sensitivity maps $\bm{\mathsf{s}}_{\ell}^{(i-1)}$ yields the linear R2D2 image update defined in \cref{Eq:image_estimate}. Conversely, when updating the sensitivity maps, fixing the image estimate $\bm{x}^{(i-1)}$ renders the forward model linear with respect to $\bm{\mathsf{s}}_{\ell}$. Specifically, $\bm{\mathsf{F}}_{\Omega}(\bm{\mathsf{s}}{_\ell} \odot \bm{x}^{(i-1)})
= \bm{\mathsf{F}}_{\Omega}(\bm{x}^{(i-1)} \odot \bm{\mathsf{s}}_{\ell})$,
which motivates the following sensitivity estimation subproblem:
\begin{equation}
\arg \min_{\bm{\mathsf{s}}_{\ell}} \frac{1}{2}\|\bm y_{\ell}-\bm{\mathsf{\bar \Phi}}^{(i-1)} \bm{\mathsf{s}}_{\ell} \|^2_2,
\label{Eq:sens_opt}
\end{equation}
where $\bm{\mathsf{\bar \Phi}}^{(i-1)} \triangleq \bm{\mathsf{F}}_{\Omega} \bm{x}^{(i-1)}$. Instead of solving \cref{Eq:sens_opt} explicitly, iR2D2 calculates the back-projected sensitivity data residual:
\begin{equation}
\bm{\bar{r}}_{\ell}^{(i-1)} = \bar{\kappa}^{(i-1)}{{\bm{\mathsf{\bar\Phi}}^{\dagger}}^{(i-1)}}\bm{\mathsf{D}}\bm{\mathsf{M}}\left({\bm{y}_{\ell}}- {\bm{\mathsf{\bar \Phi}}}^{(i-1)}  \bm{\mathsf{s}}^{(i-1)}_{\ell}\right),
\label{Eq:sens_residual}
\end{equation}
where the normalization constant is dynamically updated as
$\bar{\kappa}^{(i-1)}= [\max (| {{\bm{\mathsf{\bar{\Phi}}}^{\dagger}}^{(i-1)}}\bm{\mathsf{D}}\bm{\mathsf{M}} \bm{\mathsf{ \bar{\Phi}}}^{(i-1)}\bm{\delta}|)]^{-1}$. Here, $\bm{\mathsf{M}}$ denotes a frequency mask matrix designed to enforce the spectral smoothness inherent to coil sensitivity profiles by retaining only the central 20\% of k-space data. This effectively suppresses high-frequency components in the residual, thereby reducing contamination from anatomical structures and noise. 

Leveraging this residual, the framework implements a learned residual refinement strategy using a secondary series of DNN modules $\{\bm{\mathsf{\bar{G}}}_{\bm{\widehat{\theta_{\rm s}}^{(i)}}}\}_{1\leq i \leq I}$ akin to the image update rule defined in \cref{Eq:image_estimate}. At iteration $i$, the sensitivity map of each coil is updated and normalized as:
\begin{equation}
    {\bm{\mathsf{s}}}_{\ell}^{(i)} = \alpha^{(i)}[{\bm{\mathsf{s}}}_{\ell}^{(i-1)} + \bm{\mathsf{\bar{G}}}_{\bm{\widehat{\theta}_{\rm s}^{(i)}}}(\bm{\bar{r}}_{\ell}^{(i-1)}, {\bm{\mathsf{s}}}_{\ell}^{(i-1)})], 
    \label{Eq:sens_estimate}
\end{equation}
where $\alpha^{(i)}$ is a normalization factor defined as: 
\begin{equation}
    \alpha^{(i)}=\left(\sum_{\ell=1}^L |{\bm{\mathsf{s}}}_{\ell}^{(i-1)} + \bm{\mathsf{\bar{G}}}_{\bm{\widehat{\theta}_{\rm s}^{(i)}}}(\bm{\bar{r}}_{\ell}^{(i-1)}, {\bm{\mathsf{s}}}_{\ell}^{(i-1)})|^2\right)^{-1/2}.
    \label{Eq:alpha}
\end{equation} 
This normalization ensures strict consistency with the physical forward model in \cref{Eq:multicoil_MRI} and concentrates the signal energy within the reconstructed image. 

An illustration of the iR2D2 algorithm is shown in \cref{fig:jcr2d2}. It highlights the coupled evolution of image and sensitivity estimates. While the image reconstruction series progressively resolves fine anatomical details from initial smooth structures, the alternating sensitivity refinement steps dynamically correct coil profiles to match the improved image estimates, ensuring data consistency throughout the iterations. At the first iteration, $\bm{x}^{(0)}=\bm{0}$, implies $\bm{\mathsf{\bar \Phi}}^{(0)}=\bm{0}$ and thus $\bm{\bar{r}}_{\ell}^{(0)}=\bm{0}$. 
The initial sensitivity maps $\bm{\mathsf{s}}_{\ell}^{(0)}$ are provided by ESPIRiT to ensure a physically plausible and stable initialization. Similar to \cref{Eq:final_image_estimate}, the final sensitivity estimate for coil $\ell$ after $I$ iterations is
\begin{equation}
    \bm{\mathsf{\widehat{s}}}_{\ell} \triangleq {\bm{\mathsf{s}}}_{\ell}^{(I)} = \sum_{i=1}^I \bm{\mathsf{\bar{G}}}_{\bm{\widehat{\theta}_{\rm s}^{(i)}}}(\bm{\bar{r}}_{\ell}^{(i-1)}, {\bm{\mathsf{s}}}_{\ell}^{(i-1)}). \label{Eq:final_sens_estimate}
\end{equation}

At each iteration, the updated sensitivity maps are used to recompute the image-domain operators as \(\bm{\mathsf{\Phi}}_{\ell}^{(i-1)} \triangleq \bm{\mathsf{F}}_{\Omega}\,\bm{\mathsf{s}}_{\ell}^{(i)}\).
Accordingly, the input residual is computed by the image Data Residual (DR) layer as: %which aggregates the data fidelity term across all coils:
\begin{equation}
\bm{r}^{(i-1)} = \kappa^{(i-1)} \sum_{\ell=1}^{L}{{\bm{\mathsf{\Phi}}_{\ell}^{\dagger}}^{(i-1)}}\bm{\mathsf{D}} \left(\bm{y}_{\ell}
-\bm{\mathsf{\Phi}}_{\ell}^{(i-1)} \bm{x}^{(i-1)}\right),
\label{Eq:residual_jcr2d2}
\end{equation}
where $\kappa^{(i-1)}= [\max (|\sum_{\ell=1}^{L} {{\bm{\mathsf{\Phi}}^{\dagger}}^{(i-1)}_{\ell}} \bm{\mathsf{D}} \bm{\mathsf{ \Phi}}^{(i-1)}_{\ell}\bm{\delta}|)]^{-1}$.
The first residual can be computed using the first updated sensitivity maps $\{\bm{\mathsf{s}}_{\ell}^{(1)}\}_{\ell=1}^{L}$ as
\(
\bm{r}^{(0)} = \bm{x}_{\rm b}^{(0)}
= \kappa^{(0)} \sum_{\ell=1}^{L} {\bm{\mathsf{\Phi}}_{\ell}^{\dagger}}^{(0)}\bm{\mathsf{D}} \,\bm{y}_{\ell}.
\)
The image update then follows \cref{Eq:image_estimate}, and the final reconstruction is given by \cref{Eq:final_image_estimate}. 

\subsection{Update Conditions}
\label{sec:convergence}

Alternating optimization algorithms, such as BCFB, and standard unrolled networks both allow for sequential updates between variables (\emph{e.g.},~image and sensitivity maps). However, they typically rely on a fixed, predefined schedule, executing a fixed number of iterations per block regardless of the intermediate error landscape.
While iR2D2 is reminiscent of these schemes, it fundamentally generalizes them. By enforcing a sufficient energy descent condition independently across both update paths, the framework removes the need to predefine a fixed sequence. Instead, it dynamically dictates the number of updates allocated to each block. This adaptive formulation guarantees meaningful convergence and strictly non-increasing residual errors across the iterations, preventing updates driven by numerical noise.

At each iteration $i > 1$, a relative improvement threshold is enforced via this condition, beginning with the sensitivity update step. 
Following the sensitivity map update, $\bm{\mathsf{s}}^{(i)}$ is accepted only if the aggregate residual energy across all coils decreases by a minimum relative margin compared to the previous state.
We introduce an empirical tolerance factor $\zeta = 0.99$, chosen from validation set convergence behavior. It enforces a strict 1\% minimum decrease in energy, filtering out marginal, noise-driven updates that do not meaningfully improve the optimization.
\begin{equation}
\sum_{\ell=1}^{L} \|\bar{\kappa}^{(i-1)}{{\bm{\mathsf{\bar\Phi}}^{\dagger}}^{(i-1)}}\bm{\mathsf{D}}\bm{\mathsf{M}}({\bm{y}_{\ell}}- {\bm{\mathsf{\bar \Phi}}}^{(i-1)}  \bm{\mathsf{s}}^{(i)}_{\ell})\|^2_2 < \zeta\sum_{\ell=1}^{L} \| \bm{\bar{r}}_{\ell}^{(i-1)} \|_2^2.
\label{eq:uacbar}
\end{equation}
If the DNN prediction fails to satisfy this energy decrease threshold, the update is rejected; the forward pass bypasses the current module's output, actively maintaining the previous sensitivity estimate $\bm{\mathsf{s}}^{(i-1)}_{\ell}$ for the subsequent operations.

This identical conditional criterion is applied independently to the image reconstruction update via UC.
The image update $\bm{x}^{(i)}$ is accepted only if:
\begin{equation}
    \| \kappa^{(i-1)} \sum_{\ell=1}^{L}{{\bm{\mathsf{\Phi}}_{\ell}^{\dagger}}^{(i-1)}}\bm{\mathsf{D}} (\bm{y}_{\ell}
-\bm{\mathsf{\Phi}}_{\ell}^{(i-1)} \bm{x}^{(i)})
\|_2^2< \zeta\| \bm{r}^{(i-1)} \|_2^2 .
\label{eq:uac}
\end{equation}
Similarly, if this condition fails, the network rejects $\bm{x}^{(i)}$ and propagates the previous image $\bm{x}^{(i-1)}$ forward.

The complete step-by-step execution of iR2D2 is detailed in \cref{alg:iR2D2}. Since $\overline{\text{UC}}$ and UC are evaluated sequentially but independently, any iteration can result in one of four distinct states: (1) both updates are accepted; (2) the sensitivity map is refined but the image update is rejected; (3) the sensitivity map is rejected but the image is refined using the preserved map; or (4) both updates are rejected, passing the previous stable state forward. As a result, the number of consecutive updates for each variable is determined dynamically.
\begin{algorithm}
\small
\caption{iR2D2 algorithm}\label{alg:iR2D2}
\begin{algorithmic}
\Require $(\bm{\mathsf{s}}_{\ell}^{(0)})_{1\leq\ell\leq L}, \bm{x}^{(0)}=0$, total iteration $I$ (See \cref{convergence})
\For{$i = 1$ \textbf{to} $I$}

% Sensitivity map update
\Comment{Sensitivity self-calibration update}
\State $\bm{\mathsf{\bar \Phi}}^{(i-1)} \gets \bm{\mathsf{F}}_{\Omega} \bm{x}^{(i-1)}$
\State $\bm{\bar{r}}_{\ell}^{(i-1)} \gets \text{Compute according to \cref{Eq:sens_residual}, } \forall \ell=1,\ldots, L$
\If{Training}
    \State $\bm{\widehat{\theta}_{\rm s}}^{(i)} \gets \text{Train according to \cref{Eq:loss_JCR2D2}}$
\EndIf    
\State $\bm{\mathsf{s}}_{\ell}^{(i)} \gets \alpha^{(i)}[{\bm{\mathsf{s}}}_{\ell}^{(i-1)} + \bm{\mathsf{\bar{G}}}_{\bm{\widehat{\theta}_{\rm s}^{(i)}}}(\bm{\bar{r}}_{\ell}^{(i-1)}, {\bm{\mathsf{s}}}_{\ell}^{(i-1)})], \quad \forall \ell=1,\ldots, L$
\If{$\overline{\text{UC}}$ in \cref{eq:uacbar} not satisfied}
    \State $\bm{\mathsf{s}}_{\ell}^{(i)} \gets \bm{\mathsf{s}}_{\ell}^{(i-1)}, \quad \forall \ell=1,\ldots, L$
\EndIf

% Image update
\Comment{Image reconstruction update}
\State $\bm{\mathsf{\Phi}}^{(i-1)}_{\ell} \gets \bm{\mathsf{F}}_{\Omega}\bm{\mathsf{s}}_{\ell}^{(i)}, \quad \forall \ell=1,\ldots, L$
\State $\bm{r}^{(i-1)} \gets \text{Compute according to \cref{Eq:residual_jcr2d2}}$
\If{Training}
    \State $\bm{\widehat{\theta}}^{(i)} \gets \text{Train according to \cref{Eq:loss_R2D2}}$
\EndIf    
\State $\bm{x}^{(i)} \gets \bm{x}^{(i-1)} + \bm{\mathsf{G}}_{\bm{\widehat{\theta}^{(i)}}}(\bm{r}^{(i-1)}, \bm{x}^{(i-1)})$
\If{UC in \cref{eq:uac} not satisfied}
    \State $\bm{x}^{(i)}\gets \bm{x}^{(i-1)}$
\EndIf
 
\EndFor
\end{algorithmic}
\end{algorithm}

\subsection{Normalization}
\label{normalisation}
To mitigate generalization issues caused by discrepancies in pixel value ranges between the training data and the targets (\emph{e.g.},~test images), normalization techniques must be applied. We implement iteration-specific normalization during both the training and inference phases, treating the image reconstruction and the sensitivity estimation step independently.  

During the training, for the image reconstruction modules, at any iteration $i > 1$, the input dataset is normalized by dividing each image triplet $(\bm{x}_k^{\star}, \bm{x}_k^{(i-1)}, \bm{r}_k^{(i-1)})$ by $\beta^{(i-1)}_{k}$, which is the mean value of the previous image magnitude estimate $\bm{x}^{(i-1)}_{k}$. The first DNN module is also trained using a normalized dataset, where each image pair $(\bm{x}_k^{\star}, \bm{r}^{(0)}_k)$ is divided by $\beta^{(0)}_{k}$, the mean pixel magnitude of the back-projected residual $\bm{r}^{(0)}_k$, since $\bm{x}^{(0)} = \bm{0}$. During the inference (see \cref{Eq:image_estimate,Eq:final_image_estimate}), all DNN modules are applied to normalized inputs, with their outputs denormalized accordingly. This is implemented via the transformations
\(\bm{\mathsf{G}} \mapsto \beta \bm{\mathsf{G}}(\cdot / \beta)\), where $\beta \in \mathbb{R}_{\ge 0}$
is the iteration-specific scalar derived for the image reconstruction step. 

In contrast, the sensitivity estimation modules do not require a global scaling factor, as coil sensitivity profiles are inherently bounded between $0$ and $1$. 
The initial estimates ${\bm{\mathsf{s}}_{k}^{(0)}}$ derived from ESPIRiT, as well as the target maps $\bm{\mathsf{s}}^{\star}_{k}$, inherently satisfy the constraint defined in \cref{Eq:sens}.

\subsection{DNN series training} 
To minimize discrepancies between the estimated and GT images, the networks are trained sequentially in an alternating fashion. First, the sensitivity estimation DNN is trained to refine the coil profiles. The loss function for the $i$-th sensitivity network is defined as a residual-learning $\ell_1$-norm objective:
\begin{equation}
\begin{aligned}
\underset{\bm{\theta}^{(i)}_{\rm s}}{\min} \; & \frac{1}{K}\frac{1}{L}
\sum_{k=1}^{K}\sum_{\ell=1}^{L}
\| (\bm{\mathsf{s}}^{\star}_{\ell})_{k} \\
& - \alpha^{(i)}\big[
(\bm{\mathsf{s}}^{(i-1)}_{\ell})_{k}
+ \bm{\mathsf{\bar{G}}}_{\bm{\theta}_{\rm s}^{(i)}}(
(\bm{\bar{r}}^{(i-1)}_{\ell})_{k},
(\bm{\mathsf{s}}^{(i-1)}_{\ell})_{k})
\big] \|_1
\end{aligned}
\label{Eq:loss_JCR2D2}
\end{equation}
where $K$ is the number of training samples and $L$ is the number of coils. The GT sensitivity maps $\bm{\mathsf{s}}^{\star}$ were generated using the Birdcage coil sensitivity simulation implemented in SigPy \cite{sigpy}. By including normalization $\alpha^{(i)}$ introduced in \cref{Eq:alpha} within the backpropagation chain, the training loss forces the network to learn an update that remains optimal after the unity constraint is applied.

Subsequently, the image reconstruction module is trained using the updated sensitivity maps (implicitly incorporated via the residual calculation in \cref{Eq:residual_jcr2d2}). The loss function for the $i$-th image network is defined as:
\begin{equation}
	\underset{\bm{\theta}^{(i)}}{\min} \frac{1}{K}\sum_{k=1}^{K}\| \bm{x^{\star}}_{k} -[\bm{x}^{(i-1)}_{k} + \bm{\mathsf{G}}_{\bm{\theta}^{(i)}}(\bm{r}^{(i-1)}_{k}, \bm{x}^{(i-1)}_{k})] \|_1,
	\label{Eq:loss_R2D2}
\end{equation}

The output of each trained DNN is employed to iteratively update the image estimates, as specified in \cref{Eq:image_estimate}, and the corresponding data residual as per \cref{Eq:residual}. These updated image pairs are subsequently included in the training dataset for the next DNN in the sequence.  

To accelerate convergence, we employ a progressive initialization for both updates. The learnable parameters of the first network ($i=1$) are initialized using the Kaiming (He) \cite{he2015delving} uniform distribution. For all subsequent iterations ($i \geq 2$), the networks are initialized using the weights from the previously trained module, allowing the series to focus on refining high-frequency residuals rather than learning from scratch.

\subsection{Number of DNN modules and series convergence} \label{convergence}

Formally, the number of DNN modules $I$ is an algorithmic parameter, and an appropriate procedure is required to determine its optimal value. The sequential training process for the R2D2 and iR2D2 is assumed to be terminated when the evaluation metrics for the reconstruction quality computed on a validation dataset reach stabilization. The $I$ DNN modules trained are then available for reconstruction.

\subsection{R2D2-Net}
\label{R2D2-Net}
An unrolled incarnation of R2D2, named R2D2-Net \cite{aghabiglou2024r2d2,chen2024scalable}, can be developed for benchmarking purposes by unrolling the iterative R2D2, with a predetermined number of internal iterations $I$. 
Similar to NC-PDNet, this configuration transforms the iterative process into an end-to-end unrolled DNN where the image reconstruction modules utilize distinct parameters at each stage. Both unrolled architectures use a single DNN module $\bm{\mathsf{\bar{G}}}_{\widehat{\bm \theta}_{\rm s}}$ to refine the sensitivity maps before unrolling the image iterations. The resulting updated measurement operator is then fixed and embedded within the DR layers across all subsequent image reconstruction iterations. Like NC-PDNet, R2D2-Net is characterized by substantial computational complexity and significant memory requirements during training. This limitation arises from the inclusion of exact NUFFT operations within the DR layers, which are embedded in the backpropagation graph across all $I$ iterative steps. Consequently, the model exhibits reduced scalability and computational efficiency compared to R2D2 and iR2D2. Despite these structural similarities, R2D2-Net fundamentally differs from NC-PDNet in its core algorithmic structure. While NC-PDNet relies on cross-iteration memory buffers to propagate state information and utilizes a different underlying DNN architecture, R2D2-Net employs the specific Matching Pursuit-based structure of R2D2, where the learnable modules are strictly tasked with estimating additive residual image components.

\subsubsection{Unrolled Cascade}
The forward pass of R2D2-Net begins by refining the initial sensitivity maps using the DNN module. The updated maps are generated and normalized as:
\begin{equation}
    \bm{\widehat{\mathsf{s}}}_{\ell}=\gamma\left(\bm{\mathsf{\bar{G}}}_{\widehat{\bm \theta}_{\rm s}}(\bm{\mathsf{s}}_{\ell}^{(0)})\right),
\end{equation}
where the normalization $\gamma=(\sum_{\ell=1}^L |\bm{\mathsf{\bar{G}}}_{\bm{\widehat{\theta}_{\rm s}}}({\bm{\mathsf{s}}}_{\ell}^{(0)})|^2)^{-1/2}$. Similar to the $\alpha$ parameter introduced for iR2D2, this normalization enforces that the estimated sensitivity maps strictly satisfy the constraint specified in \cref{Eq:sens}.

Subsequently, the unrolled image reconstruction cascade begins, utilizing the forward measurement operator for each coil updated to $\bm{\widetilde {\mathsf{\Phi}}}_{\ell}\triangleq \bm{\mathsf{F}}_{\Omega} \bm{\widehat{\mathsf{s}}}_{\ell}$ using refined sensitivity maps. Each subnetwork $\bm{\mathsf{G}}_{\bm{\widehat{\theta}^{(i)}}}$ with $1 \leq i \leq I$ takes two images: the previous output $\bm{x}^{(i-1)}$ and its associated back-projected data residual $\bm{r}^{(i-1)}$ as the inputs. 
This residual is computed using the updated measurement operator:
\begin{equation}
    \bm{r}^{(i-1)}=\kappa \sum_{\ell=1}^{L} \bm{\widetilde{\mathsf{\Phi}}}_{\ell}^{\dagger}\bm{\mathsf{D}}\left(\bm{y}_{\ell} - \bm{\widetilde{\mathsf{\Phi}}}_{\ell} \bm{x}^{(i-1)}\right).
\end{equation}

\subsubsection{Training}
In R2D2-Net, the image reconstruction modules and single sensitivity refinement module are trained simultaneously. All learnable parameters are optimized jointly to minimize the reconstruction error of the final image estimate. The joint loss function is defined as:
\begin{equation}
    \underset{\bm{\theta}^{(1)},\dots, \bm{\theta}^{(I)},\bm{\theta}_{\rm s}}{\min} \frac{1}{K}\sum_{k=1}^{K}\| \bm{x^{\star}}_k - 
    \sum_{i = 1}^{I}(\bm{\mathsf{G}}_{\bm{\theta}^{(i)}}(\bm{r}^{(i-1)}_k, \bm{x}^{(i-1)}_{k})) \|_1.
	\label{Eq:loss_R2D2Net}
\end{equation}
The dependence on the sensitivity parameters $\bm{\theta}_{\rm s}$ is established implicitly through the calculation of the data residual $\bm{r}^{(i-1)}_k$ using the updated measurement operator $\bm{\widetilde {\mathsf{\Phi}}}_{\ell}$, which is directly formed from the refined maps $\bm{\widehat{\mathsf{s}}}_{\ell}$.

\subsubsection{Normalization}

At the training stage, R2D2-Net is trained using a normalized dataset, where each image pair $(\bm{x}_k^{\star}, (\bm{x}_{{\rm b}})_k)$ is divided by $\beta^{(0)}_{k}$. Within the subnetworks, the intermediate variables $\bm{x}^{(i)}_{k}$ and the back-projected data residuals $\bm{r}^{(i)}_{k}$ with $1< i <I$ remain unnormalized. Instead, we rely on end-to-end training to learn the appropriate internal normalization. At the inference stage, R2D2-Net is applied to normalized inputs, with its outputs denormalized accordingly. 

%%%%%%%%%%%%%%%%%%%%%%%%%%%%%%%%%%%%%%%%%%
\section{Training details}\label{sec:simulation}
\vspace{-2pt}

\subsection{Network architectures} 
\subsubsection{U-Net}
Following from R2D2 developments in radio astronomy, the first DNN architecture utilized within R2D2 is U-Net \cite{ronneberger2015u}. Within the proposed iR2D2 framework and the benchmark R2D2-Net, this architecture is specifically deployed as the sensitivity map estimation module ($\bm{\mathsf{\bar{G}}}$) to robustly capture smooth spatial coil profiles. The network comprises symmetric compression and expansion paths. It employs four pooling layers using 2D average-pooling (stride 2) for downsampling, while upsampling is achieved via 2D transposed convolutions. Skip connections are utilized to propagate high-resolution spatial features from the compression to the expansion phases, and a final $1 \times 1$ convolution layer produces the two-channel output.

\subsubsection{U-WDSR}
U-WDSR is an advanced architecture proposed in \cite{aghabiglou2025r2d2} as a high-performance alternative to U-Net for the image reconstruction modules ($\bm{\mathsf{G}}$) across the R2D2 incarnations evaluated in this study. It integrates the WDSR residual body \cite{yu2018wide}, originally designed for image and video super-resolution, into the core structural components of a U-Net backbone, leveraging wide activations and dense connections to maximize information flow. While the comprehensive architectural details can be found in \cite{aghabiglou2025r2d2}, we note that a minor variation was implemented: we deactivate weight initialization within the WDSR residual body to facilitate convergence.

\subsection{GT dataset}
Our GT dataset uses complex-valued, fully sampled images from the FastMRI knee dataset \cite{zbontar2018fastmri}. To enhance image quality and control noise levels, the raw images are preprocessed using a denoising DNN \cite{zhang2023practical}, applied separately to the real and imaginary components.
As illustrated in \cref{fig:dataset}(a), the denoised images are resized from 320$\times$320 to 192$\times$192 for computational efficiency and intensity normalized to a range of $[0, 1]$. The final dataset is partitioned into 24,915 training and 4,233 validation images.

\subsection{Data generation details} \label{sec:data_generation}

\begin{figure}
  \centering
\includegraphics[width=1\linewidth]{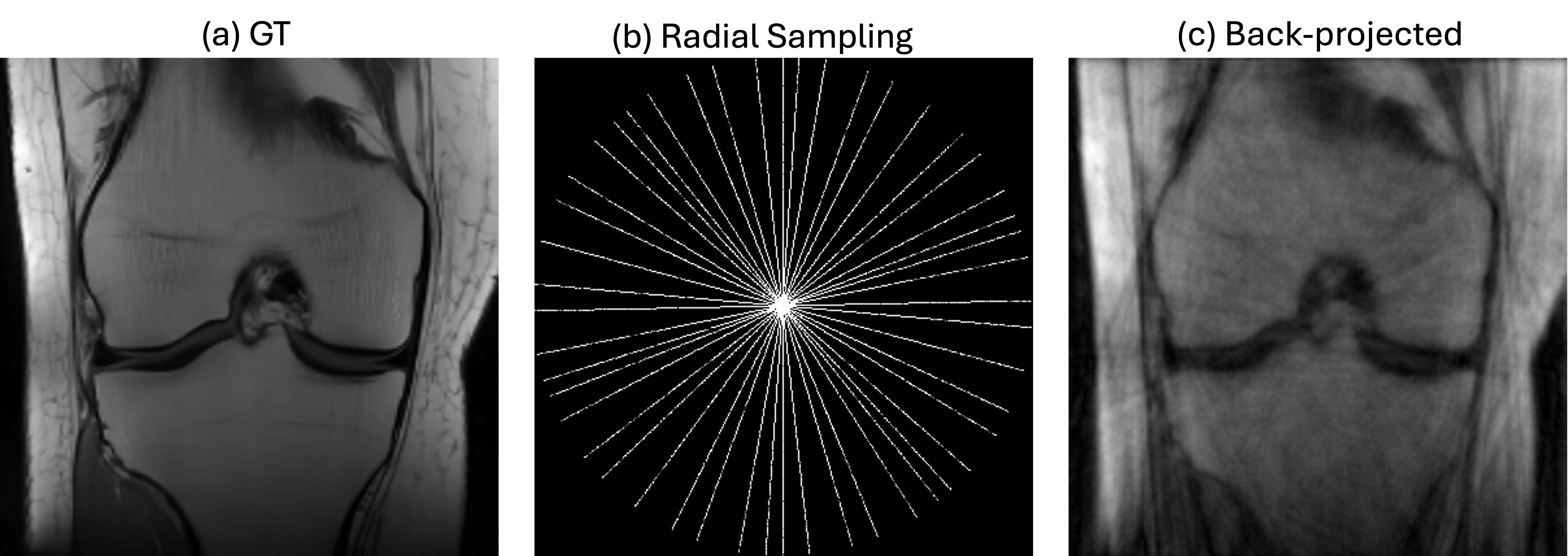}

  \caption{Illustration of the non-Cartesian MRI problem. Panel (a) displays an example GT image of size $192\times192$. 
  Panel (b) illustrates the radial sampling pattern in the spatial Fourier domain with $\text{AF}=8$ (24 spokes containing 9,216 samples), superimposed to the Fourier transform of the GT image. Panel (c) shows the corresponding back-projected image.}
  \label{fig:dataset}
  \vspace{-5mm}
\end{figure}

We simulate accelerated 2D radial k-space trajectories where each radial spoke contains $N_{\rm p} = 2\sqrt{N}$ sampling points uniformly distributed across the k-space diameter. To ensure an efficient and relatively uniform angular coverage, the spokes are generated using the small golden angle increment of 68.25° \cite{wright2014non}. The sparsity of the k-space sampling is determined by the total number of spokes, $N_{\text{s}}$. Following \cite{ramzi2022nc}, we define the Acceleration Factor (AF) for 2D radial sampling as ${\rm AF} = \sqrt{N}/N_{\text{s}}$. While previous studies typically train models for a single, fixed AF (\emph{e.g.},~4 or 8), we treat $N_{\text{s}}$ as a randomized variable during model training. This strategy forces the network to develop a generalized reconstruction capability adaptable to varying acceleration regimes. An example of the resulting 2D sampling trajectory at AF = 8 is illustrated in \cref{fig:dataset}(b).
We randomly select the number of spokes $N_{\text{s}} \in \{10,\dots, 79, 80\}$, corresponding to AFs ranging from 19.2 to 2.6, and the number of coils $L \in \{8,\dots, 31, 32\}$, to generate the non-Cartesian radial trajectories. 

The k-space measurements are then simulated using these trajectories and the GT images based on the forward model in \cref{Eq:multicoil_MRI}, with the addition of complex Gaussian noise. 
To ensure robust training, the noise level is carefully calibrated so that the target dynamic range of the reconstruction matches the inherent dynamic range of the GT image. For our normalized images, the GT dynamic range is defined as the reciprocal of the faintest structural feature intensity, denoted by $\sigma$. To maintain a well-controlled, Gaussian-like intensity distribution across the dataset, we empirically define this nominal $\sigma$ as the 6th percentile of each specific image's magnitude.
Following \cite{wilber2023scalable}, we mathematically link this target image-domain dynamic range ($\sigma^{-1}$) to the additive k-space noise standard deviation $\tau_{\ell}$ via the relation $\tau_{\ell} = \sigma\sqrt{2{\rm L}_{\ell}^2/{\rm L}_{\ell}^{'}}$, where ${\rm L}_{\ell}$ and ${\rm L}_{\ell}^{'}$ are the spectral norm of the measurement operator for the $\ell$-th coil when the density compensation weighting is applied once and twice, respectively.  
For each GT image, based on a randomly selected radial sampling pattern, we generate one back-projected image (\emph{e.g.},~\cref{fig:dataset}(c)), creating a pair of samples used for supervised learning. %Thus, the numbers of inverse problems for training and validation are 24915 and 4233, respectively.

\subsection{Benchmark algorithms}
The performance of iR2D2 is evaluated against several established benchmarks, categorized as follows: (i) Pure DNNs U-Net and  U-WDSR, which recover images directly from the initial back-projected data without enforcing physics-based data consistency; (ii) Physics-guided solvers, including NC-PDNet (an unrolled architecture specifically tailored for non-Cartesian MRI) and DDS (a PnP-style approach interlacing diffusion models with data fidelity updates); and (iii) Prior R2D2 incarnations, specifically the baseline R2D2 algorithm and its end-to-end unrolled variant R2D2-Net, both evaluated utilizing U-Net and U-WDSR as their core network backbones.

\subsection{Implementation details}
Following \cite{ramzi2022nc}, all DNN implementations treat the real and imaginary parts of complex-valued tensors as two separate channels to utilize vanilla convolutions. Consequently, across all R2D2 incarnations (R2D2, R2D2-Net, and iR2D2), every core DNN module accepts a 4-channel input tensor. For the initial module, this tensor concatenates a zero-initialized complex estimate with the initial complex-valued back-projected data. All subsequent modules process a concatenation of the previously refined complex estimate and the current back-projected data residual.

Regarding network capacities, the image reconstruction modules based on U-Net utilize 64 channels in their initial convolution layer. In contrast, U-WDSR-based reconstruction modules are restricted to 32 initial channels to accommodate GPU memory limitations during unrolled training. In iR2D2 framework, the sensitivity map estimation modules employ a U-Net architecture initialized with 32 channels. In both R2D2 and iR2D2, the initial sensitivity map estimation $\bm {\mathsf{s}}^{(0)}$ is estimated by ESPIRiT. The sequential training process was terminated at $I=10$ and $I=14$ for R2D2 and iR2D2, respectively. The number of subnetworks for both R2D2-Net (U-Net) and R2D2-Net (U-WDSR) are set to $10$. This choice of parameters was made to optimise performance. 

In the NC-PDNet implementation, the buffer strategy from the original paper \cite{ramzi2022nc} is retained with a buffer size of 5, and a 3-layer CNN with 32 convolution filters is employed. The number of unrolled iterations is set to 10.

For DDS, we use $100$ sampling steps with the stochasticity parameter as $0.85$ following \cite{song2020denoising}. The number of conjugate gradient steps is 5, and the regularization parameter of DDS is fine-tuned through grid search to optimize the reconstruction quality, with respect to the noise level of the target MR image.

In NC-PDNet, both training and inference use inputs normalized by $\beta^{(0)}_{k}$. 
The deviation from the normalization strategy proposed in \cite{ramzi2022nc} is motivated by the improved reconstruction quality achieved with the proposed normalization approach.

All algorithms were executed on a single NVIDIA A100-SXM4-40GB GPU on the DiRAC Tursa system at the University of Edinburgh, using the deep learning library PyTorch \cite{paszke2019pytorch}. The training batch sizes for R2D2, iR2D2, NC-PDNet, and R2D2-Nets were 4, 4, 4, and 1, respectively. The above models were trained using the Adam optimizer \cite{kingma2014adam} with a learning rate of $0.0001$. For DDS, we followed \cite{dhariwal2021diffusion} and trained the model for one million iterations with the AdamW optimizer \cite{loshchilov2017fixing} and a learning rate of $0.0001$. The NUFFT implementation is TorchKbNufft \cite{muckley20}.

%%%%%%%%%%%%%%%%%%%%%%%%%%%%%%%%%%%%%%%%%%
\section{Validation on simulated data} \label{sec:sim_results}

\subsection{Evaluation dataset}

We randomly selected 50 GT images from the raw FastMRI validation dataset. The testing followed the same simulation procedure as training and validation described in \ref{sec:data_generation}, with the number of spokes $N_{\text{s}}$ in $\{12,16,24,32,48,64\}$, corresponding to 6 AFs: $\{16,12,8,6,4,3\}$, and the number of coils fixed to 16. For each GT image, one inverse problem is created with each of the 6 AFs, resulting in a total of 300 problems.

\subsection{Evaluation metrics}
To evaluate global reconstruction quality, we employ the Peak Signal-to-Noise Ratio (PSNR) and the Structural Similarity Index Measure (SSIM). 
Both metrics are computed on the magnitude of the reconstructed and GT images, with higher PSNR values signifying superior reconstruction fidelity and an SSIM of one indicating perfect structural similarity.

To evaluate data consistency and convergence behavior, we introduce the image-domain Residual Data Ratio (RDR). This metric is defined generally as the $\ell_2$-norm of the current back-projected residual $\bm{r}^{(i)}$, normalized by the initial back-projected image $\bm{x}_{\mathrm{b}}$:
\begin{equation}
\text{RDR} = \frac{\|\bm{r}^{(i)}\|_2}{\|\bm{x}_{\mathrm{b}}\|_2}.
\end{equation}

Similarly, to assess the fidelity of the sensitivity map refinement, we introduce $\overline{\text{RDR}}$. Since the initial sensitivity residual is zero ($\bm{\bar{r}}^{(0)}_{\ell} = \mathbf{0}$), we normalize the aggregate sensitivity residual by the first effective update generated by iR2D2:
\begin{equation}
\overline{\text{RDR}} = \left(\frac{\sum_{\ell=1}^{L} \| \bm{\bar{r}}_{\ell}^{(i)} \|_2^2}{\sum_{\ell=1}^{L} \| \bm{\bar{r}}_{\ell}^{(1)} \|_2^2}\right)^{1/2}.
\end{equation}
Applying this fixed denominator to all benchmark methods establishes a consistent and mathematically fair basis for cross-model comparison.

\subsection{Quantitative results} \label{Quantresults}

Quantitative assessments are summarized in \cref{tab:comparison_table}, reporting PSNR, SSIM, and computational costs averaged over 300 test inverse problems. For R2D2 incarnations, metrics are reported at the iteration of convergence (as detailed in the caption). 
\begin{table*}[ht!]
\caption{Quantitative reconstruction results, reconstruction times, and scalability indicators.}
\centering
        \resizebox{\linewidth}{!}{\begin{tabular}{lccccccccc}
			\toprule
			Algorithm & PSNR (dB) & SSIM &$t_{\rm loa.}$(s) & $t_{\rm inf.}$(s)& $t_{\rm res.}$(s)& $t_{\rm tot.}$(s)&Par. (M) &Tra.~Scal.&Rec.~Scal. \\
			\midrule
            
            U-Net
            &$31.48\pm4.48$&$0.88\pm0.08$ & $0.08\pm0.02$& $0.004\pm0.006$& $-$ & $0.09\pm0.02$& $31.1$&  $\checkmark$& $\checkmark$ \\
            
            U-WDSR &$34.38\pm4.19$&$0.92\pm0.06$& $0.08\pm0.02$& $0.039\pm0.015$& $-$ & $0.10\pm0.03$& $20.9$  &  $\checkmark$ & $\checkmark$\\ 

            NC-PDNet &$34.11\pm5.01$ & $0.90\pm0.08$ &$0.03\pm0.01$& $0.284\pm0.012$&$-$&$0.32\pm0.01$& $0.16$&  $\times$ & $\checkmark$ \\
			DDS &$33.73\pm6.80$&$0.85\pm0.13$ & $0.67\pm0.27$ & $82.32\pm6.43$ & $-$ & $82.39\pm6.62$ & $83.7$& $\checkmark$  & $\times$\\
			R2D2-Net (U-Net) &$34.86 \pm4.21$&$0.94\pm0.06$ &$0.08\pm0.02$&$0.322\pm0.011$& $-$& $0.38\pm0.02$& $311.2$&  $\times$& $\checkmark$\\
			R2D2-Net (U-WDSR) &${36.46\pm5.23}$&${0.94\pm0.06}$ &$0.08\pm0.02$& $0.653\pm0.012$& $-$&$0.71\pm0.02$& $209.2$&  $\times$ & $\checkmark$\\
            
			{R2D2 (U-Net)} & ${32.71\pm4.94}$&${0.91\pm0.07} $& $0.29\pm0.04$& $0.017\pm0.006$& $0.52\pm0.05$& $0.81\pm0.09$& $155.5$& $\checkmark$ & $\checkmark$\\ 
			{R2D2 (U-WDSR)}& ${35.08\pm4.67}$&${0.93\pm0.06}$& $0.29\pm0.03$& $0.093\pm0.006$& $0.28\pm0.08$& $0.68\pm0.22$& $72.7$& $\checkmark$ & $\checkmark$\\ 

            {iR2D2 (U-Net)} & ${\underline{38.01\pm6.57}}$&$\underline{0.94\pm0.07} $& $0.49\pm0.04$& $0.243\pm0.006$& $1.52\pm0.05$& $2.31\pm0.28$& $311.2$& $\checkmark$ & $\checkmark$\\ 
			{iR2D2 (U-WDSR)}& $\bm{40.21\pm6.67}$&$\bm{0.96\pm0.05}$& $0.48\pm0.08$& $0.563\pm0.011$& $5.08\pm0.08$& $5.92\pm0.38$& $315.7$& $\checkmark$ & $\checkmark$\\ 
			\bottomrule
            \vspace{1mm}
		\end{tabular}}
        
        \parbox{\linewidth}{\footnotesize%
\textbf{Note.} The first column lists the algorithm names. The second and third columns report the PSNR and SSIM values (with standard deviations) averaged over 300 test inverse problems across AF values; these values match those in \cref{fig:ni} (a-b), with the metric computed at convergence iteration $i=5$ for R2D2 (U-Net), $i=3$ for R2D2 (U-WDSR), $i=8$ for iR2D2 (U-Net), and $i=11$ for iR2D2 (U-WDSR). The best and second-best results are highlighted in bold and underlined, respectively. Columns four to seven detail the reconstruction times, broken down into model loading ($t_{\rm loa.}$), inference ($t_{\rm inf.}$) and residual calculation ($t_{\rm res.}$); the total time ($t_{\rm tot.}$) is given in the seventh column. The eighth column lists the number of parameters for each DNN architecture. Finally, scalability indicators categorize the algorithms with regard to their training scalability (column nine) or reconstruction scalability (column ten).
}
		\label{tab:comparison_table}
        \vspace{-5mm}
\end{table*}
Regarding reconstruction quality, iR2D2, utilizing the U-WDSR DNN, achieves the highest performance across all metrics, reaching 40.21 dB PSNR. This result represents a substantial improvement over R2D2 and R2D2-Net, confirming that the correction of coil sensitivities is the decisive factor in achieving state-of-the-art fidelity. Both R2D2 (U-Net) and iR2D2 (U-Net) fall short of their U-WDSR based models, indicating that the DNN structure significantly impacts the overall performance. It is also worth noting that even the single-stage U-WDSR baseline marginally outperforms both the unrolled NC-PDNet and the diffusion-based DDS, underscoring the intrinsic robustness of the U-WDSR architecture for MRI reconstruction.

\begin{figure}[ht!]
  \centering
  \setlength\tabcolsep{1.2pt}
  \begin{tabular}{cc}
       \includegraphics[width=0.5\columnwidth]{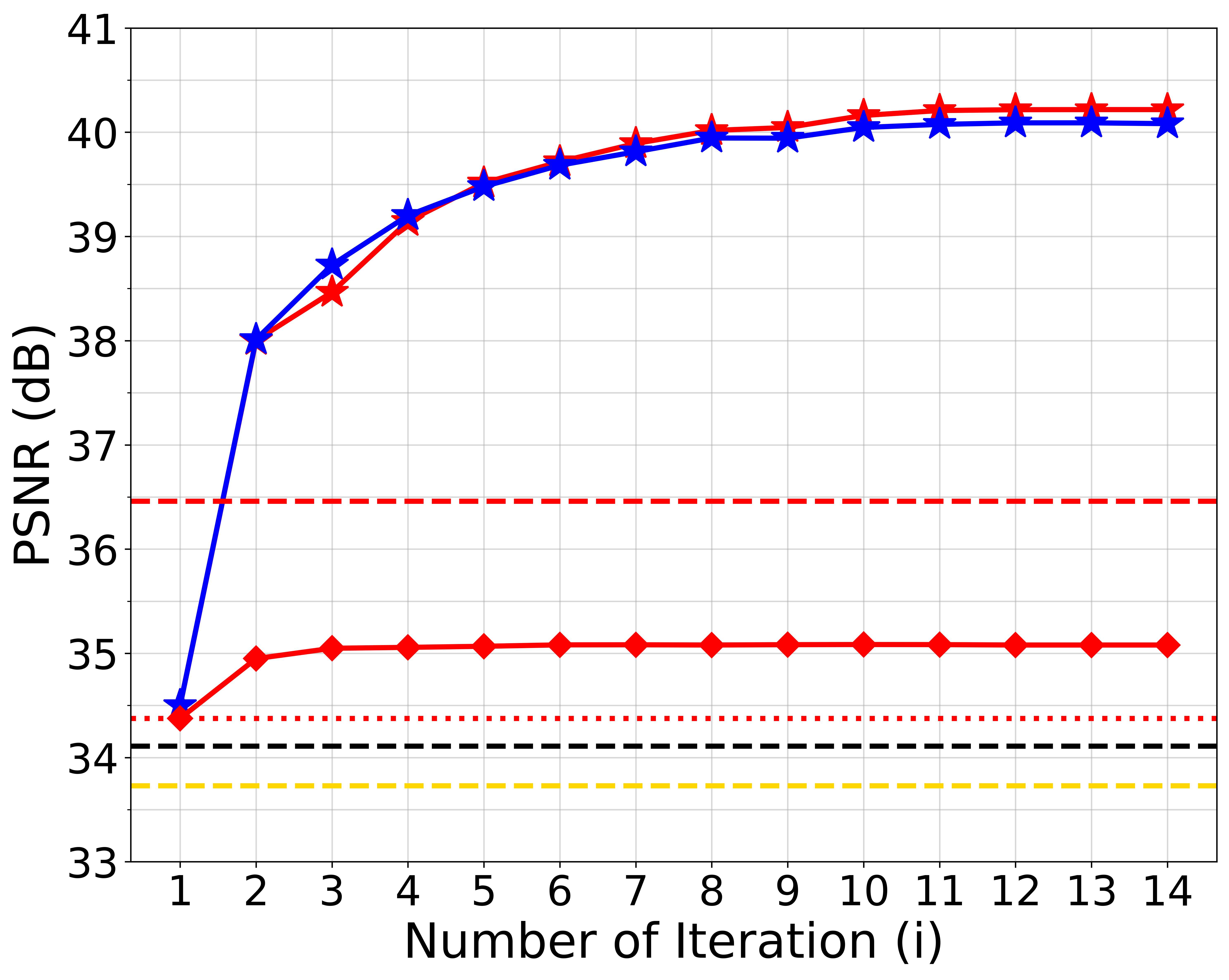}& 
    \includegraphics[width=0.5\columnwidth]{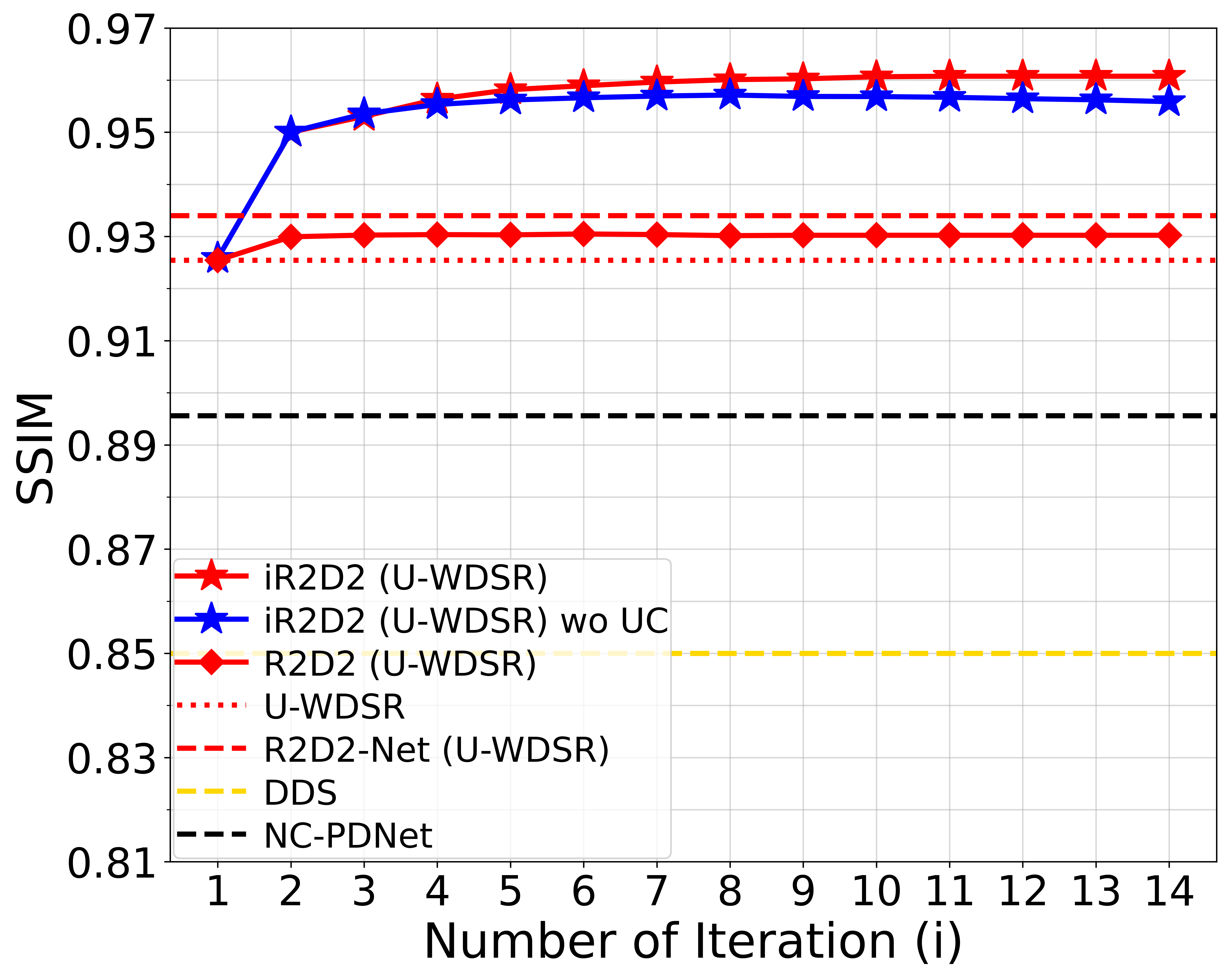}\\
    {\fontsize{8pt}{0pt}\selectfont (a) $\text{PSNR vs. iterations}$} &
       {\fontsize{8pt}{0pt}\selectfont (b) $\text{SSIM vs. iterations}$}\\
       
    \includegraphics[width=0.5\columnwidth]{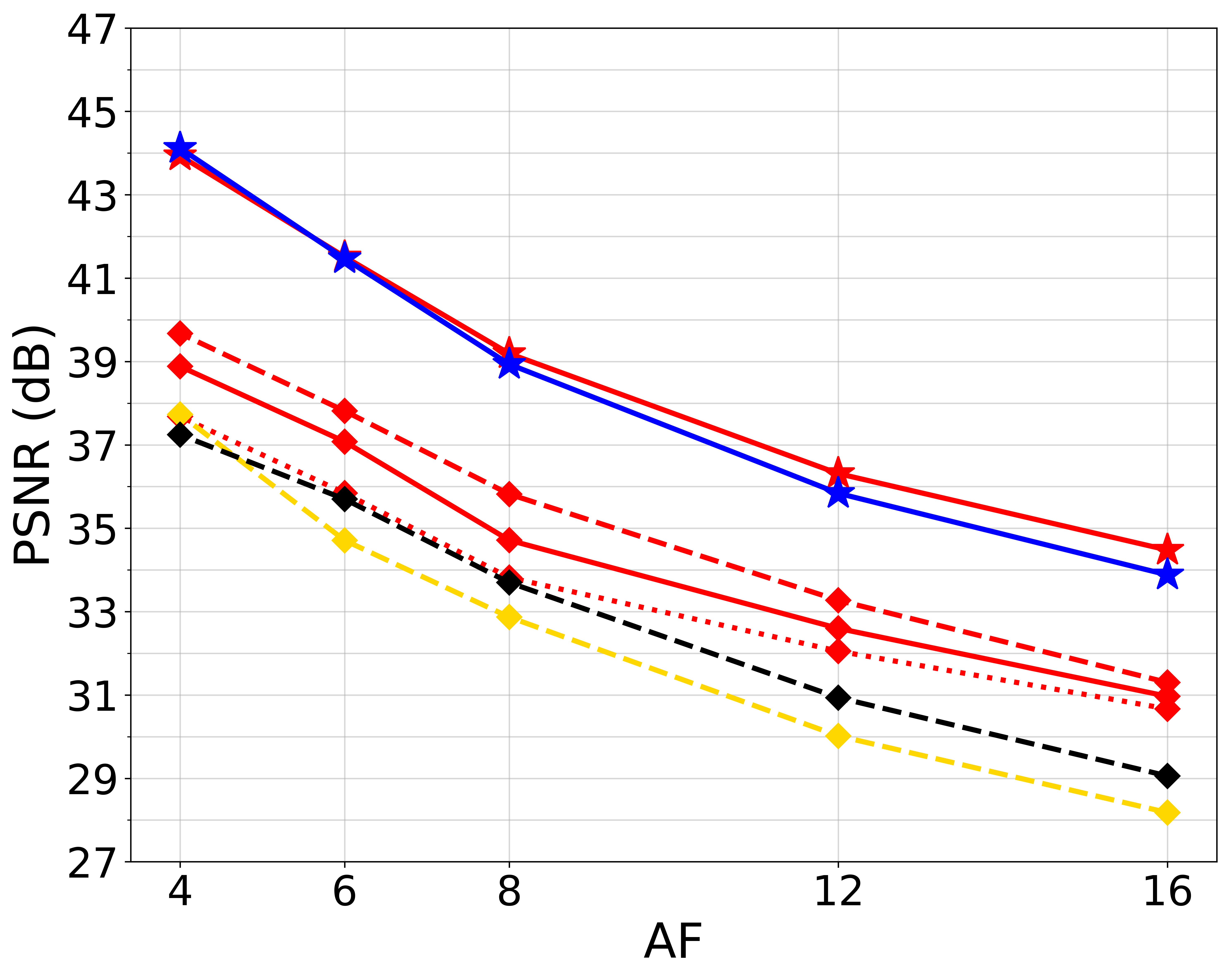}&
    \includegraphics[width=0.5\columnwidth]{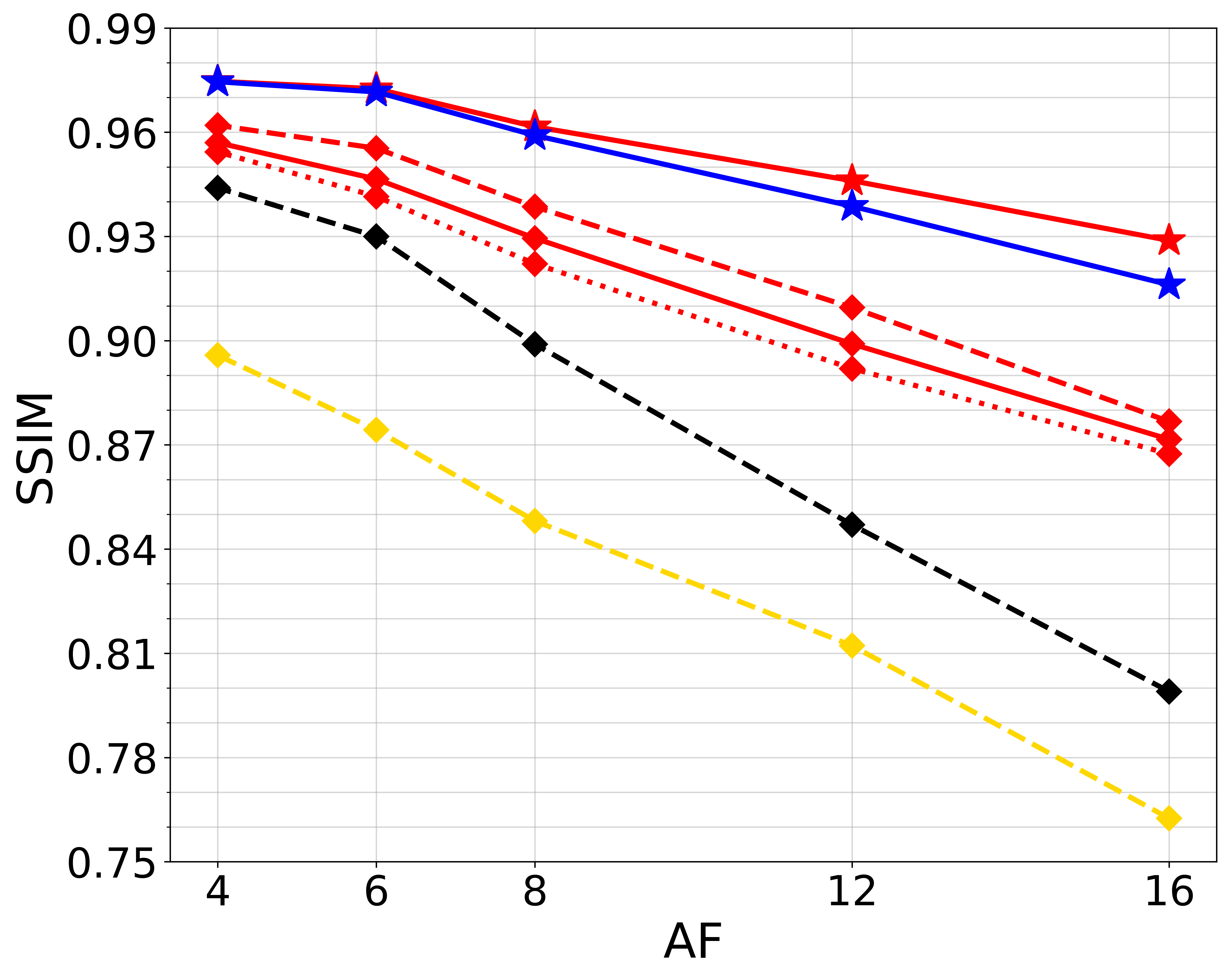}\\
    {\fontsize{8pt}{0pt}\selectfont (c) $\text{PSNR vs. AF}$}&{\fontsize{8pt}{0pt}\selectfont (d) $\text{SSIM vs. AF}$}\\

    \includegraphics[width=0.5\columnwidth]{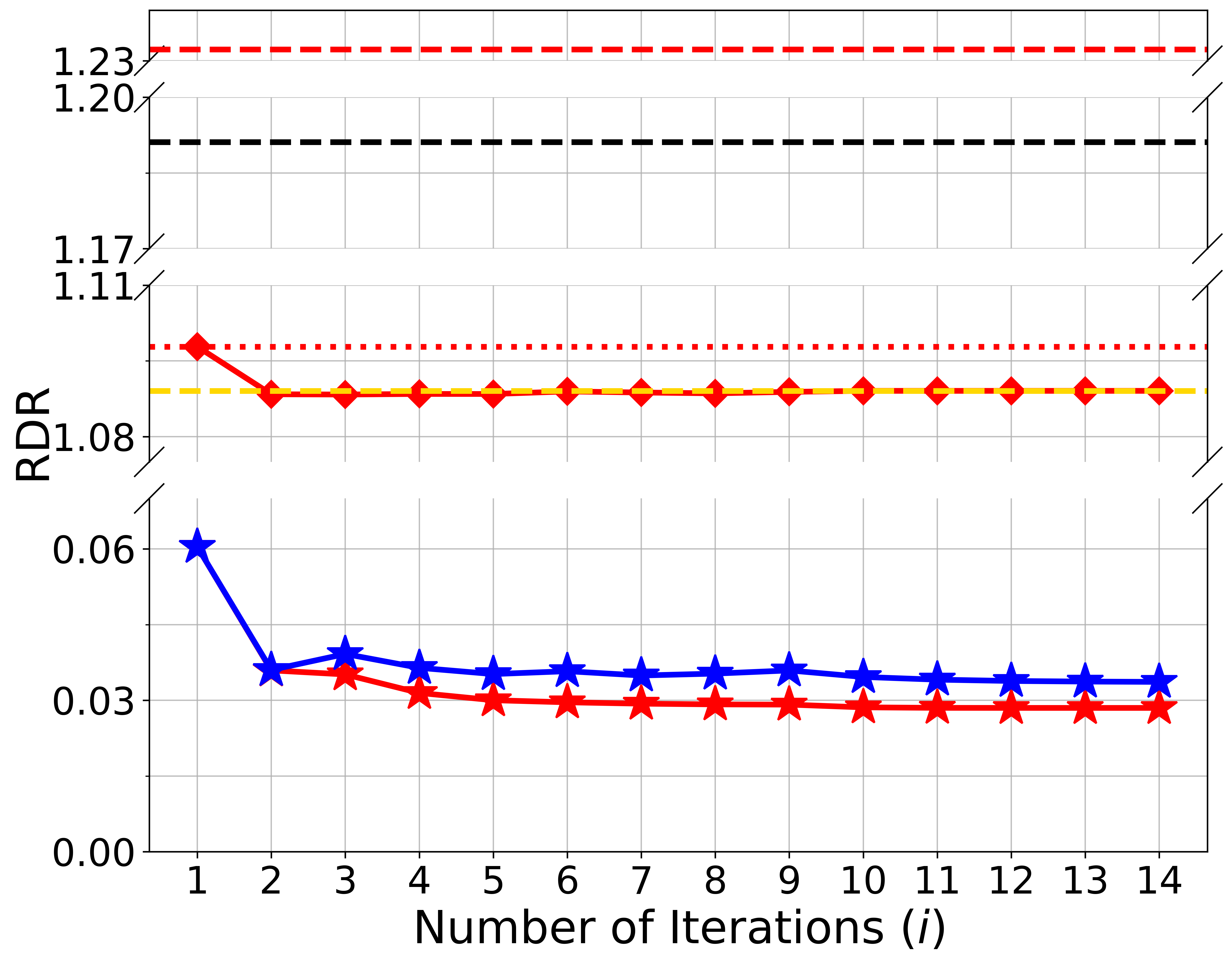}&
    \includegraphics[width=0.5\columnwidth]{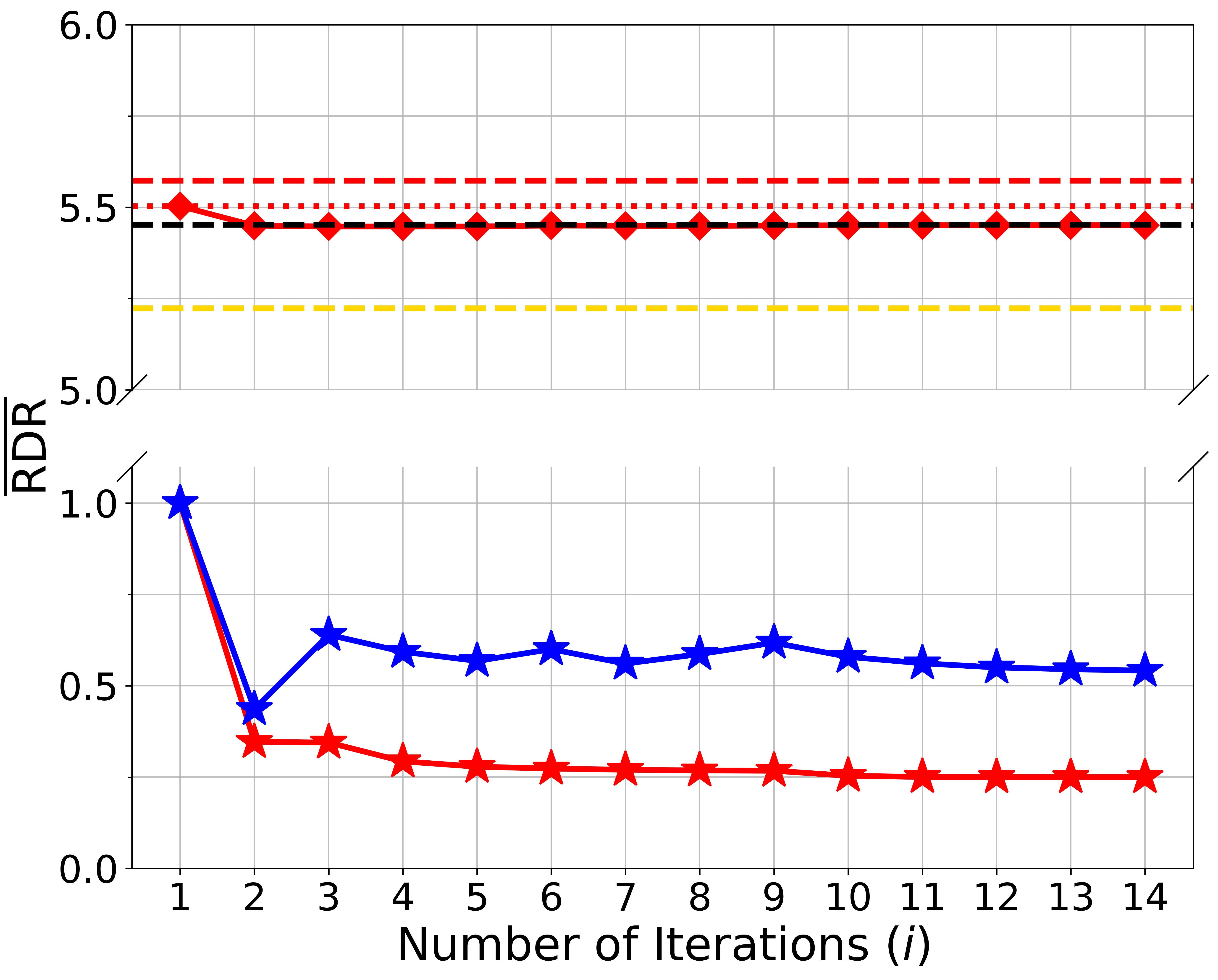}\\
    {\fontsize{8pt}{0pt}\selectfont (e) $\text{RDR vs. iterations}$}&{\fontsize{8pt}{0pt}\selectfont (f) $ \overline{\text{RDR}} \text{ vs. iterations}$}
    
  \end{tabular}
  
  \caption{Quantitative reconstruction results from simulated experiments. The performance of iR2D2 (U-WDSR) and R2D2 (U-WDSR) (solid lines) is evaluated against the number of iterations for PSNR, SSIM, RDR, and $\overline{\text{RDR}}$ in panels (a), (b), (e), and (f), respectively (reported values are averages over 300 inverse problems across all AFs). Panels (c) and (d) report the PSNR and SSIM against the AF value (reported values are averages over 50 inverse problems per AF). 
  Benchmark methods are shown as horizontal lines: U-WDSR (dotted); R2D2-Net (U-WDSR), NC-PDNet, and DDS (dashed).
  }
  \label{fig:ni}
  \vspace{-5mm}
\end{figure}

\cref{fig:ni} (a-b) compares the  PSNR and SSIM values of R2D2 (U-WDSR) and iR2D2 (U-WDSR) against the number of iterations, in comparison to benchmark algorithms. All values are averages over the 300 test inverse problems. R2D2 (U-WDSR) can be estimated to converge at iteration $i=3$, when both average PSNR and SSIM values saturate, while iR2D2 (U-WDSR) converges at iteration $i=11$. The iR2D2 (U-WDSR) model delivers substantial improvements in PSNR and SSIM compared to its corresponding DNN baseline U-WDSR. In contrast, R2D2 models that rely only on ESPIRiT sensitivity maps show marginal gains in PSNR and SSIM over the respective DNNs, likely due to inaccuracies in the residual estimation.
R2D2-Net (U-WDSR) provides satisfying performance, with PSNR and SSIM stabilizing between those of the first and second iterations of iR2D2. NC-PDNet and DDS provide significantly lower quality than iR2D2 (U-WDSR), R2D2 (U-WDSR), and R2D2-Net (U-WDSR).

\cref{fig:ni} (c-d) compares performance across the AFs. All values are averages over the 50 test inverse problems for each AF. A general decline in PSNR and SSIM is observed for all methods as AF increases. iR2D2 (U-WDSR) achieves the highest values across all AFs, demonstrating exceptional robustness. It is followed by R2D2-Net (U-WDSR), R2D2 (U-WDSR), and U-WDSR. DDS and NC-PDNet exhibit a steadier decline as AF increases.

\cref{fig:ni} (e-f) illustrates the convergence behavior and data consistency of the algorithms by tracking the RDR and $\overline{\text{RDR}}$ metrics. Both panels highlight the significant data fidelity advantage of the proposed framework. While benchmark methods fail to meaningfully reduce the relative residual energy, leaving substantial unresolved data discrepancy, iR2D2 demonstrates a steep, monotonic energy descent in both the image and sensitivity updates. This non-increasing behavior physically validates the mathematical stability enforced by the UC, proving that the self-calibration strategy successfully eliminates the calibration errors that bottleneck models relying on static sensitivity maps from ESPIRiT. Notably, the inclusion of the iR2D2 without the update conditions (blue curves) isolates the vital role of this dynamic update control. While the absence of these update conditions causes only minor degradations in the global PSNR and SSIM metrics (\cref{fig:ni}a-d), it severely destabilizes the underlying physical convergence. As explicitly shown in \cref{fig:ni}(f), without the UC and $\overline{\text{UC}}$, the sensitivity residual $\overline{\text{RDR}}$ exhibits erratic spikes and settles at a significantly higher energy state. This sharp contrast confirms that the UC and $\overline{\text{UC}}$ are not merely an empirical performance booster, but a strict mathematical necessity to prevent the amplification of numerical noise across the iterations.

\subsection{Scalability}

Table \ref{tab:comparison_table} presents the breakdown of reconstruction times averaged over 300 test problems. The results highlight a clear trade-off between speed and complexity. 
DDS exhibits the longest image reconstruction time due to its highly iterative nature (100 steps), limiting reconstruction scalability. 
Pure DNNs demonstrate the lowest image reconstruction time but at the cost of limited performance. 

The R2D2 incarnations achieve high reconstruction quality while maintaining fast reconstruction thanks to their low iteration count, \emph{i.e.}~12 iterations for iR2D2 (U-WDSR) against 3 for R2D2 (U-WDSR). While iR2D2 incurs $\sim$4 times higher computational cost compared to R2D2 due to the overhead of the alternating sensitivity estimation steps, it remains orders of magnitude faster than the diffusion-based DDS. The breakdown of reconstruction times reveals that the primary cost in iR2D2 comes from the residual calculation ($t_{\rm res}$), which involves the execution of the sensitivity modules and the application of forward/adjoint operators.
Notably, the image reconstruction process of the R2D2 incarnations spends a significant amount of time on model loading, which can be reduced by processing multiple test samples in parallel. 

\begin{figure}
  \centering
\includegraphics[width=0.5\linewidth]{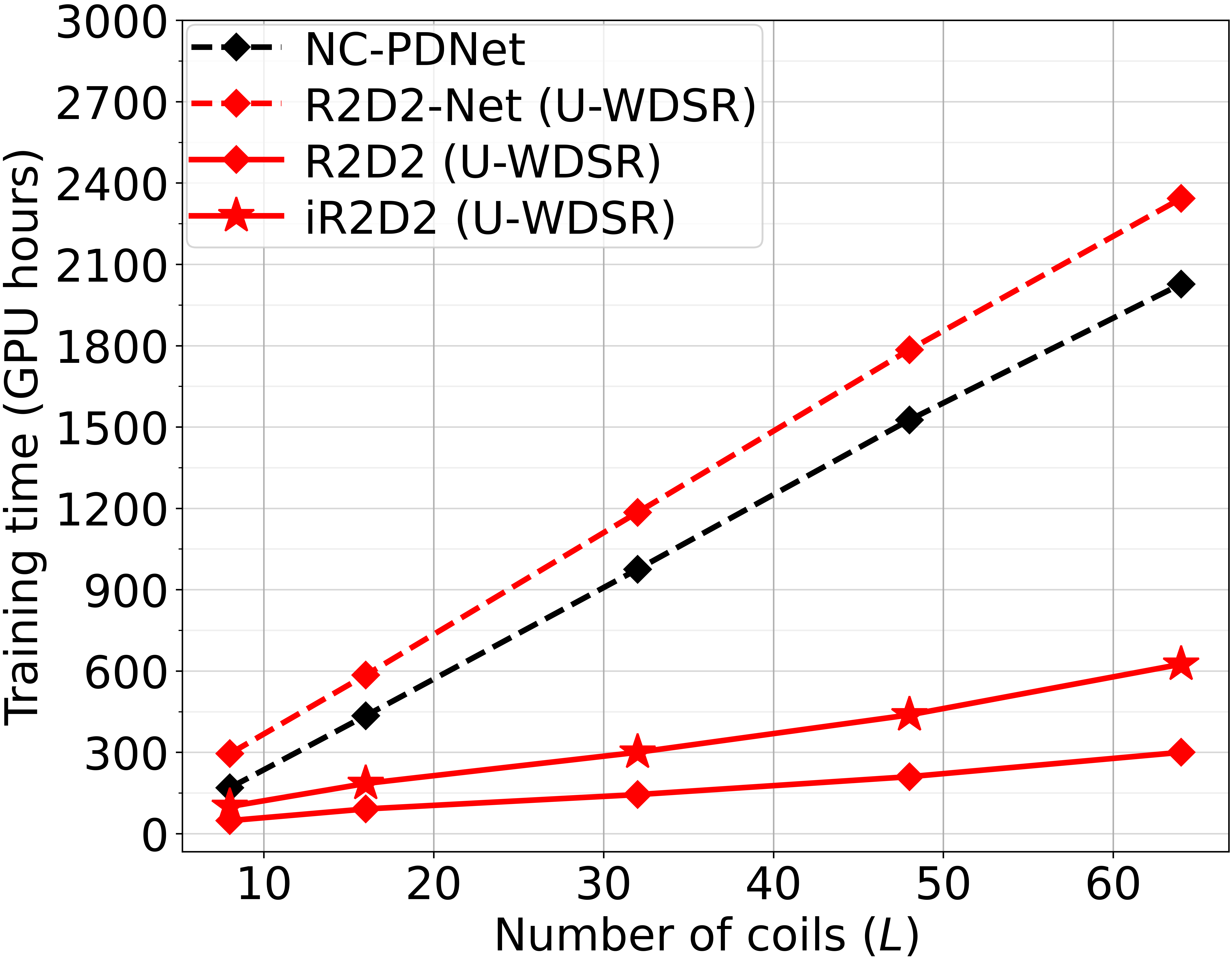}
  \caption{Training scalability comparison. Training time (GPU hours) vs. number of coils for iR2D2 (U-WDSR) (stars) and R2D2 (U-WDSR) (diamonds), shown with solid lines; NC-PDNet and R2D2-Net (U-WDSR) are shown with dashed lines (all at iteration $i=10$).}
  \label{fig:training_time}
  \vspace{-5mm}
\end{figure}

Beyond image reconstruction scalability, training scalability is also a key consideration, with training time serving as a crucial metric. 
\cref{fig:training_time} analyzes training scalability by plotting the total training duration against the coil count for four representative architectures. To ensure a consistent comparison, all models are evaluated at an identical number of iterations ($i=10$).
While all methods exhibit linear scaling, the disparity in growth rates is profound. Unrolled end-to-end architectures (NC-PDNet, R2D2-Net; dashed lines) suffer from steep increases in training costs, driven by the computational burden of backpropagating gradients through the non-Cartesian measurement operators across multiple unrolled iterations.
In contrast, R2D2 and iR2D2 (solid lines) demonstrate superior scalability with significantly shallower slopes. 
Notably, while iR2D2 incurs a moderate overhead compared to R2D2 due to the sensitivity estimation updates, it remains $\sim$4 times faster than its unrolled counterpart (R2D2-Net) at large coil counts ($L=64$). This confirms that R2D2 and iR2D2 are uniquely positioned to handle high-dimensional, many-coil acquisitions where standard unrolled methods become computationally prohibitive.
For ease of comparison, training and reconstruction scalability indicators are reported for all methods in Table \ref{tab:comparison_table}. Only U-Net, U-WDSR, R2D2, and iR2D2 scale in both training and reconstruction.

\subsection{Visual performance}

\begin{figure*}
\hspace{-6mm}
\begin{tabular}{cc}
     (a) $\mathrm{AF=4}$ \text{(48 spokes)}, 16 coils& (b) $\mathrm{AF=16}$ \text{(12 spokes)}, 16 coils \\
     \includegraphics[width=0.52\linewidth]{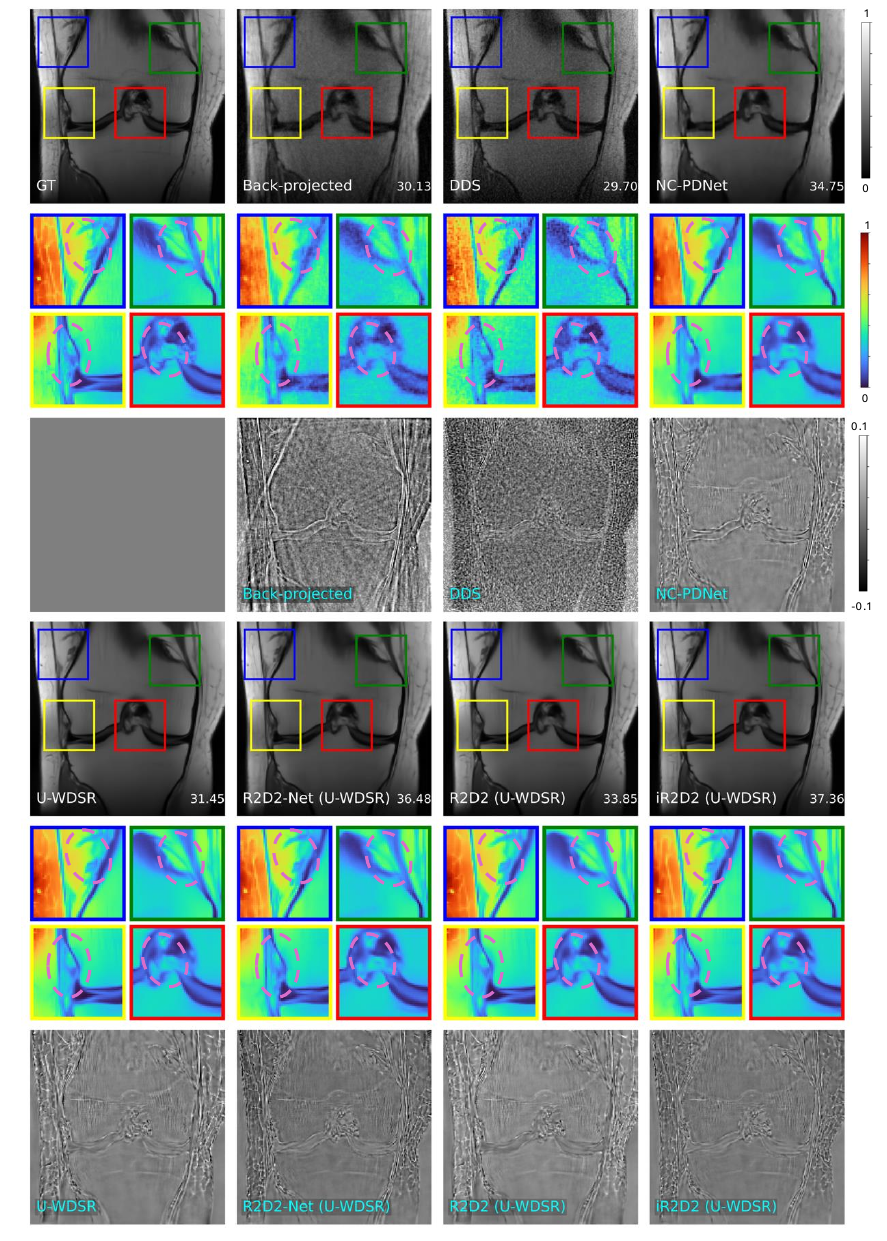}
     & \hspace{-7mm}
     \includegraphics[width=0.52\linewidth]{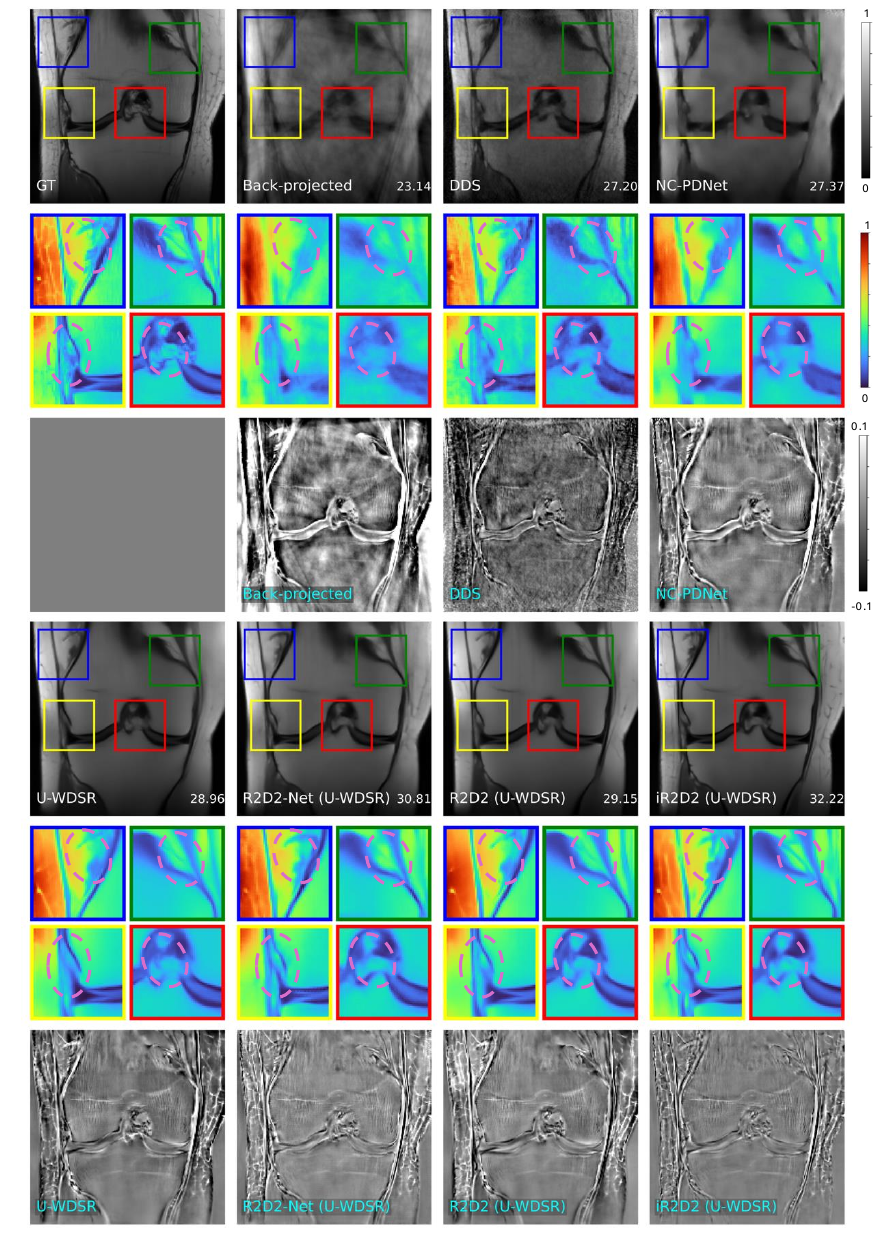}
\end{tabular}
\caption{Visual reconstruction results for one of the test inverse problems of the simulated experiments, with 16 coils for (a) $\mathrm{AF=4}$ and (b) $\mathrm{AF=16}$.
  The first and fourth rows show the GT image, back-projected, and estimated images of different methods. The second and fifth rows show the selected ROI marked by boxes in the images. Dotted ellipses in these ROIs highlight key areas with subtle differences between the estimated images. The third and sixth rows present the error images between the GT and estimated images. Values of PSNR (dB) are reported in the bottom left of the estimated images.}
\label{fig:simulation}
\vspace{-5mm}
\end{figure*}

In this subsection, we focus on the U-WDSR based methods, which already present better quantitative results, and benchmark methods. \cref{fig:simulation} compares the visual results of different methods for $\mathrm{AF=4}$ and $\mathrm{AF=16}$, respectively. 
The images displayed and corresponding PSNR values are taken at convergence iteration $i=6$ for iR2D2 (U-WDSR) and $i=3$ for R2D2 (U-WDSR). 
We firstly focus on $\mathrm{AF=16}$. The back-projected image shows significant blurring, noticeable artifacts, and substantial errors. Methods like DDS and NC-PDNet exhibit improvements in structural detail over the back-projected image but still suffer from notable reconstruction errors. The U-WDSR architecture provides a noticeable improvement in visual quality over reconstructions of DDS and NC-PDNet, similar to R2D2 (U-WDSR) and R2D2-Net (U-WDSR). iR2D2 (U-WDSR) provides the best performance, effectively preserving fine structural details and minimizing reconstruction errors. These results are in line with the quantitative analysis (see \ref{Quantresults}).

All algorithms exhibit significant improvements at $\mathrm{AF=4}$, but the general conclusions drawn at $\mathrm{AF=16}$ remain. We note that DDS produces a noisy output, with a PSNR not superior to that of the back-projected image. Interestingly, iR2D2(U-WDSR) exhibits the highest stability across accelerations, further confirming the results of the quantitative analysis. 

\subsection{Impact of dynamic range} \label{sec:dr}

\begin{figure}[ht!]
  % \centering
  \hspace{-5mm}
  \setlength\tabcolsep{1.2pt}
  \begin{tabular}{cc}
       \includegraphics[width=0.5\columnwidth]
    {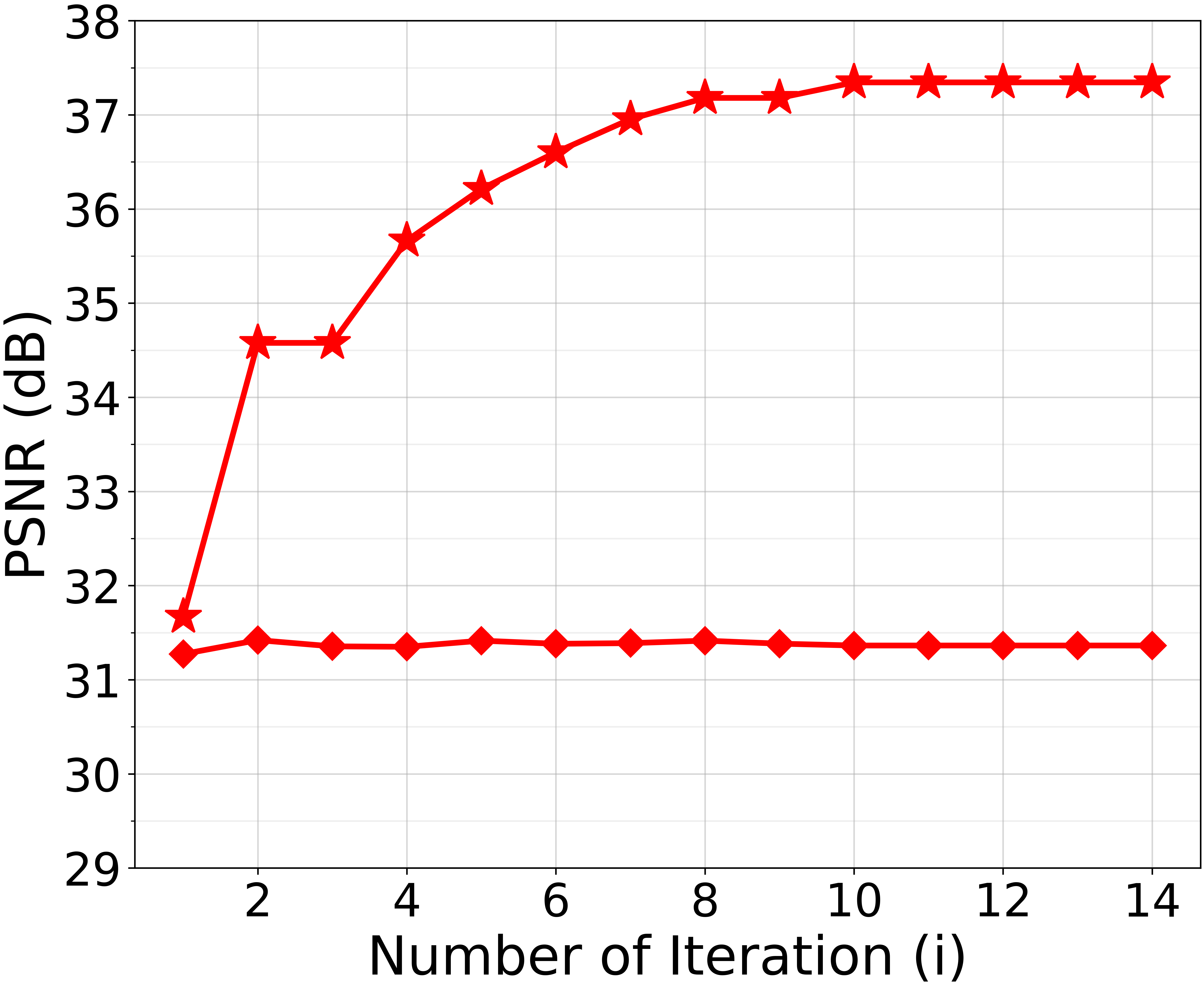}&  
    \includegraphics[width=0.5\columnwidth]{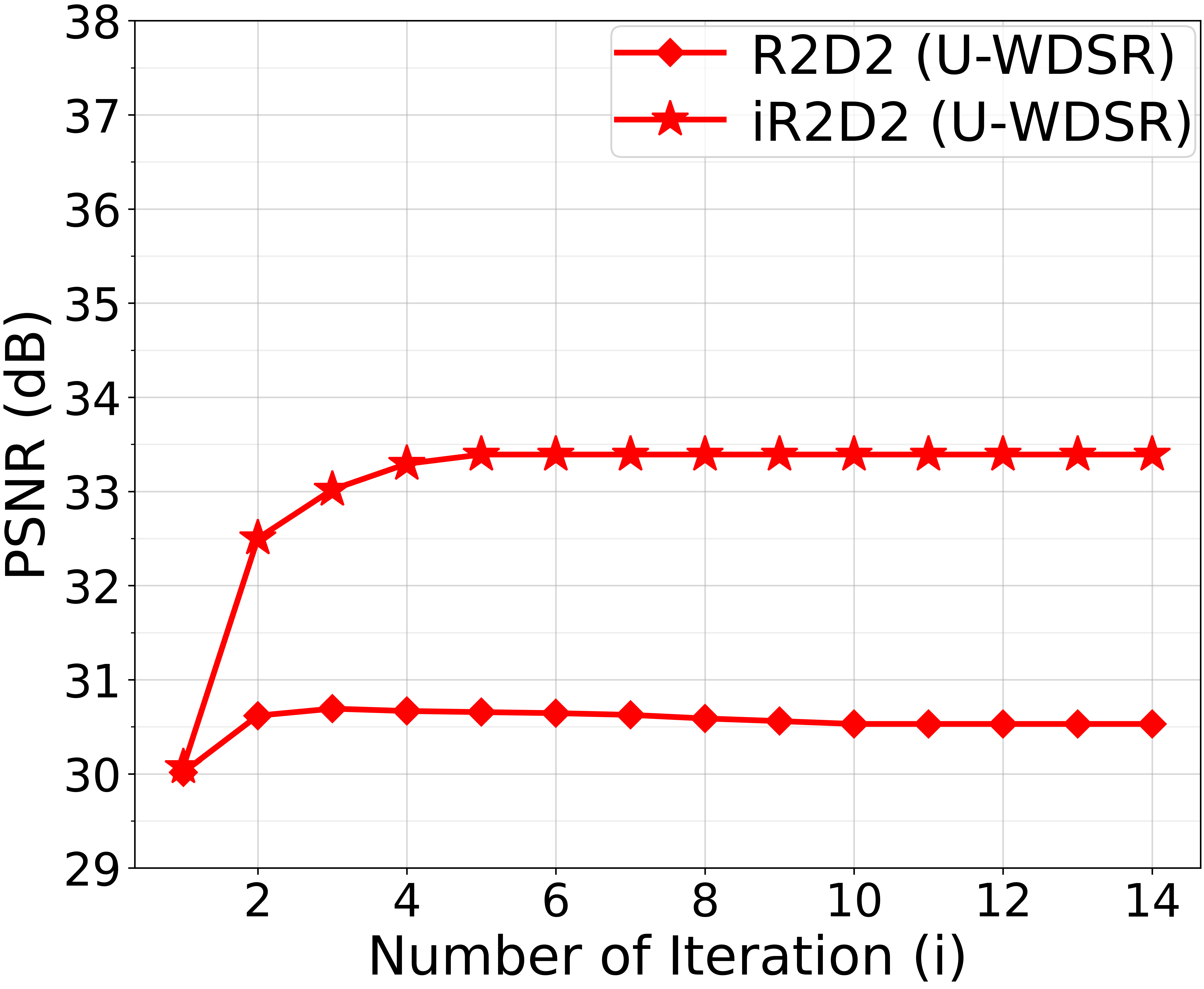}\\
       {\fontsize{8pt}{0pt}\selectfont (a) Dynamic range $=105.5$} & 
       {\fontsize{8pt}{0pt}\selectfont (b) Dynamic range $=7.7$}
  \end{tabular}
  \caption{Quantitative reconstruction results by dynamic range from simulated experiments. The PSNR evolution over iterations of R2D2 (U-WDSR) and iR2D2 (U-WDSR) is shown for one of the 300 test inverse problems at $\mathrm{AF}=8$, under high (left) and low (right) dynamic range settings.} 
  \label{fig:noise}
 \vspace{-5mm}
\end{figure}

To analyze robustness across noise regimes, we compare R2D2 (U-WDSR) and iR2D2 (U-WDSR) on randomly selected test inverse problems with specific dynamic range values in \cref{fig:noise}. We examine PSNR value variation at a high-dynamic-range case (dynamic range $=105.5$, low noise) and a low-dynamic-range case (dynamic range $=7.7$, high noise) for $\mathrm{AF=8}$.
The results reveal that the performance advantage of iR2D2 over R2D2 is most pronounced in the high-dynamic-range regime ($\sim$6 dB gap). In this clean signal environment, R2D2 saturates quickly as it hits a performance ceiling imposed by calibration errors. iR2D2, by correcting these errors, breaks through this ceiling to resolve fine details.
Conversely, at low dynamic range, the gap narrows. This corroborates the expectation that as noise becomes the dominant error source, it obscures the structural information in the residuals, limiting the potential gains from both iterative refinement and sensitivity map correction.

%%%%%%%%%%%%%%%%%%%%%%%%%%%%%%%%%%%%%%%%%%
\section{Validation on real data} \label{sec:real_results}

\subsection{Experimental setup} \label{sec:real_setting}
Data were acquired from a volunteer three years post-operative following anterior cruciate ligament (ACL) reconstruction. The subject had a permanent titanium fixation device in the femur. Imaging was performed at 3T (PrismaFit, Siemens Healthineers, Forchheim, Germany) using a 15-channel knee coil. 
Pseudo-coronal undersampled 2D golden-angle radial data were obtained with a balanced steady-state free precession (bSSFP) sequence (TE/TR = 1.58/3.32 ms, flip angle: 30-70°, slice thickness: $9$ mm, field of view: $288\times288$ mm$^2$) with a pixel size of $0.6 \times 0.6$ mm$^2$.
All human studies were approved by the local ethics committee (CER-VD approval 2021-00697), and written informed consent was given by the participant.

\subsection{Visual and quantitative performance}
 \begin{figure*}[ht!]
\hspace{-4mm}
\begin{tabular}{cc}
     (a) $\mathrm{AF=4}$ (72 spokes), 15 coils &  (b) $\mathrm{AF=8}$ (36 spokes), 15 coils\\
     \includegraphics[width=0.51\linewidth]{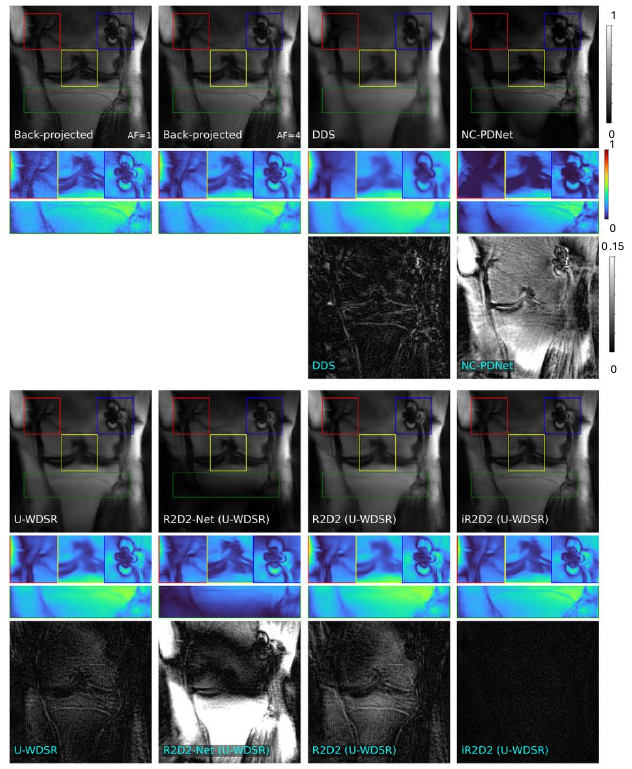}
     &\hspace{-5mm}
     \includegraphics[width=0.51\linewidth]{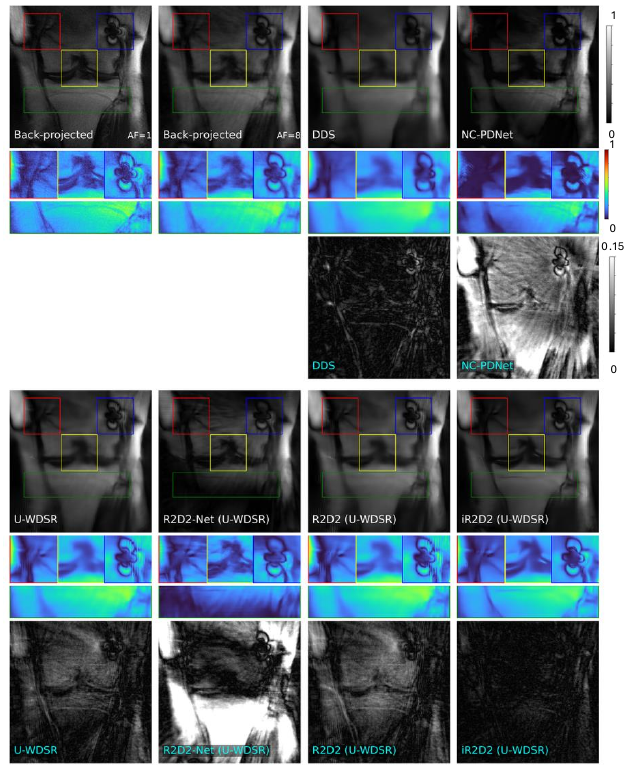}
\end{tabular}
\caption{Visual reconstruction results of real data experiment, with 15 coils acquired at (a) $\mathrm{AF=4}$ (72 spokes)  and (b) $\mathrm{AF=8}$ (36 spokes). The back-projected image with 288 sampling spokes (denoted as $\mathrm{AF=1}$) is reported for reference on the top left panel. The first and fourth rows show the back-projected images as well as reconstructions and estimated images of different methods. The second and fifth rows show the selected ROI marked by boxes in the images. The third and sixth rows present the $\bm{r}^{(I)}$.}
\label{fig:vis_real}
\vspace{-5mm}
\end{figure*}
To further validate the clinical robustness of the proposed method, \cref{fig:vis_real} presents results acquired at $\mathrm{AF=4}$ and $\mathrm{AF=8}$. We focus on four specific Regions of Interest (ROIs) to facilitate detailed analysis: the lateral femoral condyle (red), the intercondylar notch (yellow), the medial femoral condyle featuring the titanium fixation device (blue), and the proximal tibia with the physeal scar (green). We also display the final data residuals $\bm{r}^{(I)}$ to analyze the information discarded by each model.
Visual inspection reveals distinct failure modes. Both unrolled architectures (R2D2-Net, NC-PDNet) introduce severe shadow artifacts across the femoral condyles and the fixation device (red/blue ROIs). This failure is corroborated by their residual maps, which contain significant anatomical structure, indicating that the precomputed calibration priors failed to fit the data in these regions. The diffusion-based DDS avoids these shadows but suffers from characteristic over-smoothing, washing out the fine trabecular textures near the intercondylar notch (yellow ROI) and blurring the tibial physeal scar (green ROI).

R2D2 (U-WDSR) significantly improves upon the unrolled baselines, eliminating the gross shadow artifacts. However, without self-calibration, it exhibits residual haziness in fine structures at higher accelerations ($\mathrm{AF=8}$). 
iR2D2 (U-WDSR) delivers the superior reconstruction, offering the sharpest delineation of the physeal scar and trabecular details across both accelerations. Notably, the iR2D2 residual maps appear as unstructured noise, unlike the structure-heavy residuals of the benchmark algorithms. This confirms that the sensitivity self-calibration update successfully corrects domain-shift errors. By dynamically enforcing a sufficient energy descent across both sensitivity and image updates, this adaptive scheme ensures the network extracts maximal anatomical information.

\begin{figure}[ht!]
  \centering

  \setlength\tabcolsep{1.2pt}
  \begin{tabular}{cc}
       \includegraphics[width=0.5\columnwidth]
    {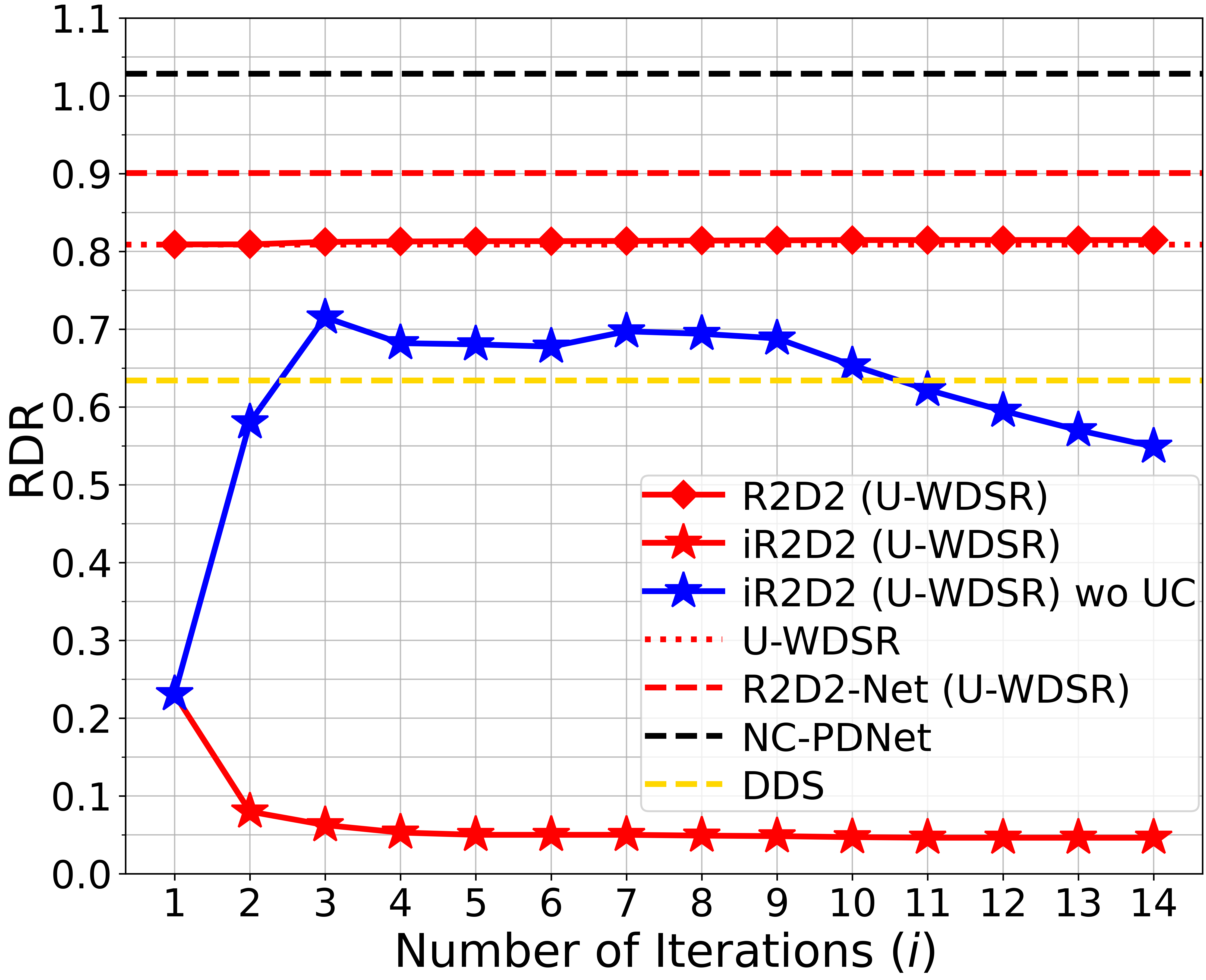}&  
    \includegraphics[width=0.5\columnwidth]{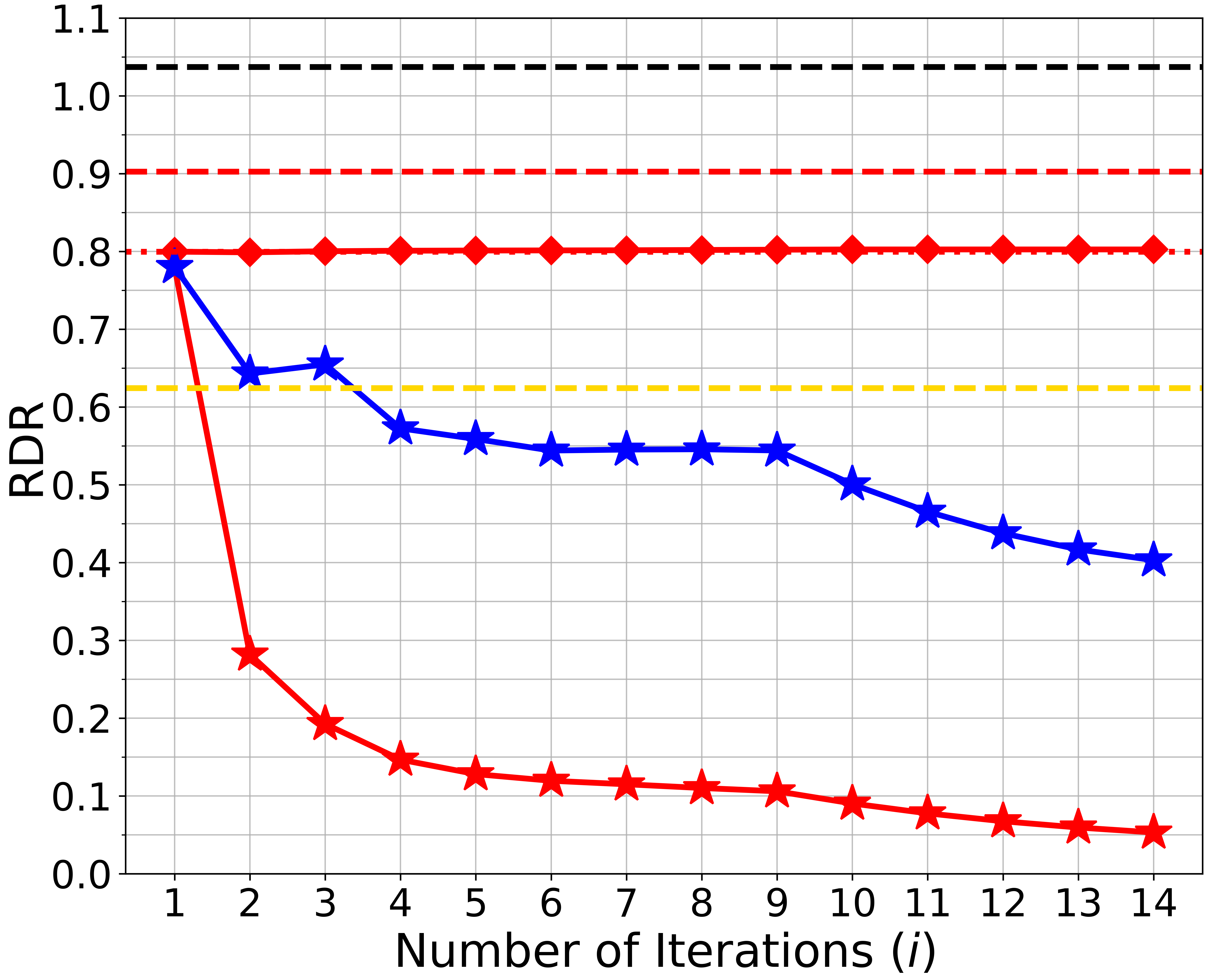}\\
       {\fontsize{8pt}{0pt}\selectfont (a) $\mathrm{AF=4}$} & 
       {\fontsize{8pt}{0pt}\selectfont (b) $\mathrm{AF=8}$}\\
       \includegraphics[width=0.5\columnwidth]
    {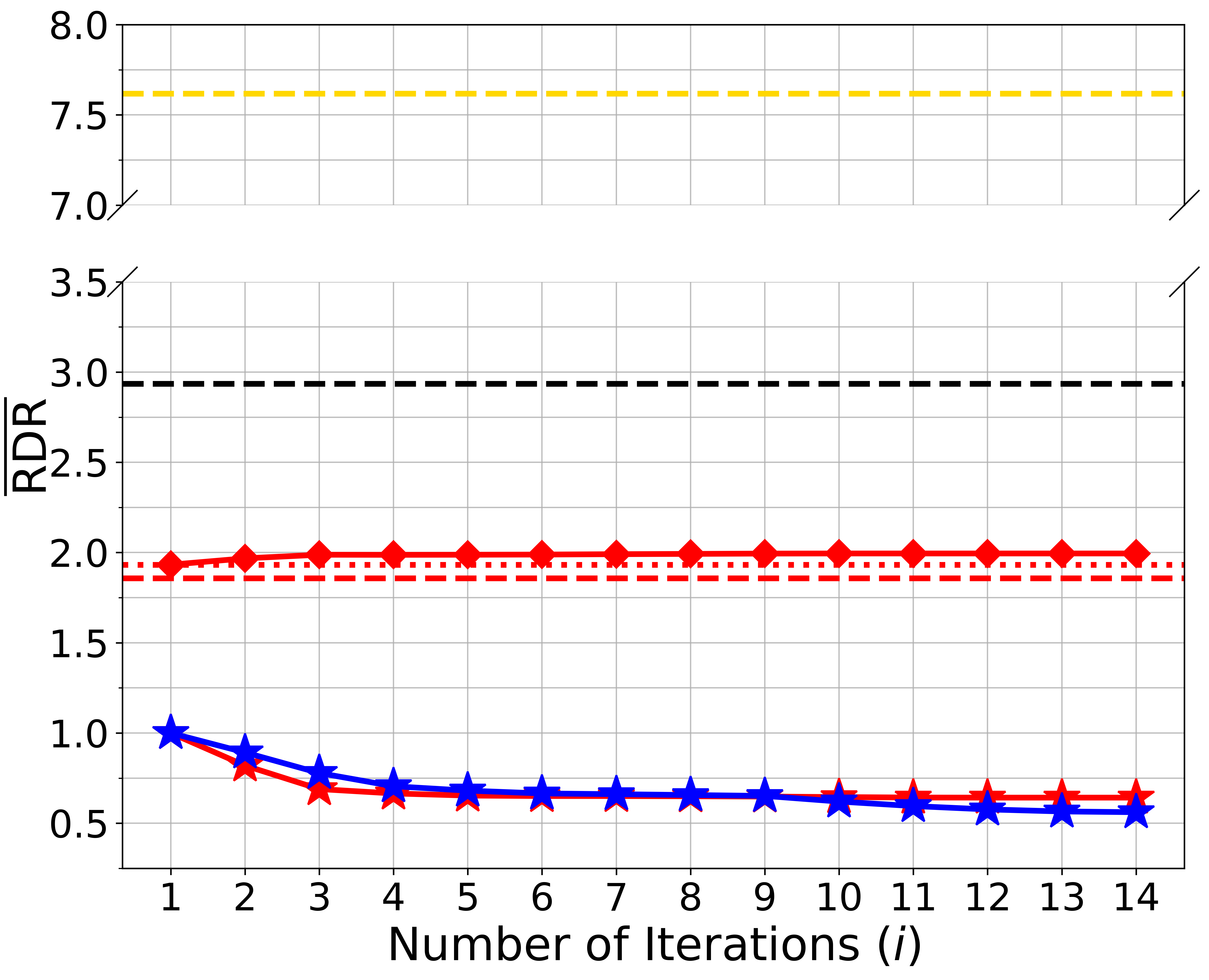}&  
    \includegraphics[width=0.5\columnwidth]{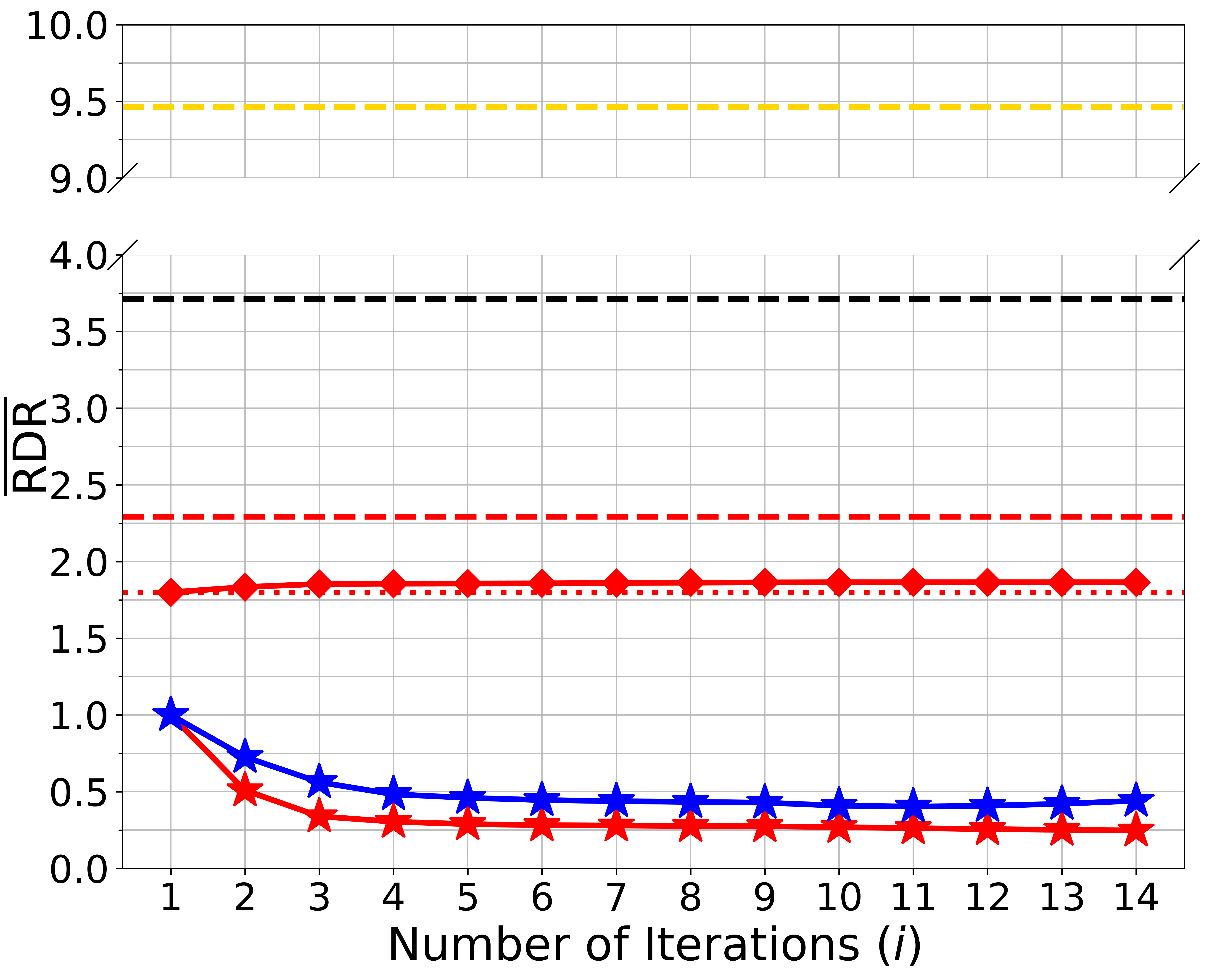}\\
       {\fontsize{8pt}{0pt}\selectfont (c) $\mathrm{AF=4}$} & 
       {\fontsize{8pt}{0pt}\selectfont (d) $\mathrm{AF=8}$}
       
  \end{tabular}

  \caption{Quantitative evaluation of data fidelity for the 15-coil real data experiment. The RDR (panels (a) and (b)) and $\overline{\text{RDR}}$ (panels (c) and (d)) for iR2D2 (U-WDSR) and R2D2 (U-WDSR) (solid lines) are plotted against the number of iterations at $\mathrm{AF}=4$ and $\mathrm{AF}=8$. Benchmark methods are evaluated on their final reconstructed outputs and represented via horizontal lines: the base U-WDSR is denoted by dotted lines, while R2D2-Net (U-WDSR), NC-PDNet, and DDS are represented by dashed lines.}
  \label{fig:sigma_real_8}
  \vspace{-5mm}
\end{figure}

Finally, in \cref{fig:sigma_real_8} we evaluate the image-domain $\text{RDR}$ and the sensitivity-domain $\overline{\text{RDR}}$ for the real data acquisitions at $\mathrm{AF}=4$ and $\mathrm{AF}=8$ . As visualized in panels (a) and (b), the baseline R2D2 exhibits a slight increase in image residual error over successive iterations, reflecting the compounding accumulation of data inconsistencies caused by static, precomputed calibration errors. In contrast, iR2D2 completely resolves this domain-shift vulnerability. Governed by the UC, the proposed framework enforces a sufficient energy descent across both acceleration factors, achieving a final $\text{RDR}$ approximately an order of magnitude lower than the benchmark methods. The indispensable role of this dynamic update condition is starkly exposed by the ablation model (blue curves). Without the UC and $\overline{\text{UC}}$, the unconstrained alternating updates become highly unstable when exposed to the complexities of real clinical acquisitions, triggering severe, erratic spikes in the image residual that fail to fully recover.
Furthermore, panels (c) and (d) confirm the stability of the self-calibration on real data. The sensitivity residual $\overline{\text{RDR}}$ drops steeply and strictly monotonically, physically validating that the refinement successfully corrects the measurement operator even under highly accelerated, real-world conditions.

%%%%%%%%%%%%%%%%%%%%%%%%%%%%%%%%%%%%%%%%%%

\section{Conclusion}\label{sec:conclusion}
In this study, we introduced iR2D2, a novel framework designed to address the reconstruction fidelity, calibration, and scalability challenges of non-Cartesian MRI. While adapting the R2D2 paradigm from radio astronomy successfully brings training scalability to MRI, it remains vulnerable to calibration errors. Specifically, precomputed sensitivity maps derived from suboptimal non-Cartesian calibration data often fail to match the acquired measurements, causing domain-shift errors. To overcome this limitation, we extent the framework into iR2D2 by integrating a secondary series of DNN modules. This bespoke interlaced architecture jointly estimates the sensitivity maps and images, enabling self-calibration that directly corrects physical model errors. Furthermore, we enhance the architecture to operate as an adaptive optimization scheme. While alternating optimization algorithms like BCFB or unrolled DNNs allow for multiple sequential updates, iR2D2 fundamentally generalizes this approach by removing the need for a predefined schedule. By integrating dynamic, error-controlled UC and $\overline{\text{UC}}$ blocks that independently enforce a sufficient residual energy descent across both updates, the framework allocates block updates dynamically based on the evolving error landscape. In addition to resolving the severe computational and memory bottlenecks of standard end-to-end unrolled networks, this algorithmic adaptability provides a fundamental advantage over their strictly constrained, static forward passes.

Extensive validation on both simulated and real data with undersampled radial k-space acquisition demonstrates that iR2D2 sets a new state of the art, outperforming established benchmarks (NC-PDNet and DDS) across all acceleration factors. These empirical validations underscore the effectiveness of the dynamic update condition, demonstrating that this adaptive scheme guarantees monotonic energy descent and mathematically stabilizes the self-calibration process even under higher acceleration.
Consequently, experiments on real data confirm that the proposed self-calibration effectively eliminates domain-shift artifacts (\emph{e.g.},~shadowing, aliasing) seen in methods relying on precomputed sensitivity priors. 
iR2D2 achieves an efficient balance between performance gains and computational cost. While sensitivity refinement adds moderate overhead compared to R2D2, the method remains orders of magnitude faster than diffusion-based solvers, reinforcing its viability for clinical deployment. Furthermore, its modular training scheme ensures scalability to high-dimensional settings where unrolled networks are computationally prohibitive. Future work will leverage this scalability to apply iR2D2 to 3D and 4D dynamic MRI, potentially incorporating advanced architectural backbones to further push the boundaries of fast, calibrated imaging.

%%%%%%%%%%%%%%%%%%%%%%%%%%%%%%%%%%%%%%%%%%%%%%%%%%
%%%%%%%%%%%%%%%%% Acknowledgments %%%%%%%%%%%%%%%%
\section*{Acknowledgements}

The authors thank JC Ye and HJ Chung for insightful discussion on DDS. The authors also thank P. Arratia for valuable comments that improved the quality of the manuscript. SC, YC, AA, MT, and YW were supported by UKRI  (grants EP/T028270/1, ST/W000970/1, APP31234, EP/S023291/1). RBvH was supported by the SNSF  (grant  CRSII5\_202276). This work used the DiRAC Tursa GPU Service at the University of Edinburgh, funded by UKRI through the STFC DiRAC HPC Facility (grant APP51638).
%%%%%%%%%%%%%%%%%%%%%%%%%%%%%%%%%%%%%%%%%%%%%%

%%%%%%%%%%%%%%%%%%%%%%%%%%%%%%%%%%%%%%%%%%%%%%
\section*{Data and code availability}
%%%%%%%%%%%%%%%%%%%%%%%%%%%%%%%%%%%%%%%%%%%%%%

Codes will be made available as part of a future release of the \href{https://basp-group.github.io/BASPLib/}{BASPLib} code library on GitHub. BASPLib is developed and maintained by the Biomedical and Astronomical Signal Processing Laboratory (\href{https://basp.site.hw.ac.uk/}{BASP}).

\bibliographystyle{IEEEtran}
\bibliography{reference}

@article{wright2014non,
  title={Non-Cartesian parallel imaging reconstruction},
  author={Wright, Katherine L and Hamilton, Jesse I and Griswold, Mark A and Gulani, Vikas and Seiberlich, Nicole},
  journal={J. Magn. Reson. Imaging},
  volume={40},
  number={5},
  pages={1022--1040},
  year={2014},
  publisher={Wiley Online Library}
}

@INPROCEEDINGS{chen2024scalable,
  author={Chen, Yiwei and Tang, Chao and Aghabiglou, Amir and Chu, Chung San and Wiaux, Yves},
  booktitle={Proc. EUSIPCO}, 
  title={Scalable Non-Cartesian Magnetic Resonance Imaging with {R2D2}}, 
  year={Aug. 2024},
  pages={1511-1515}
}

@article{glover2011overview,
  title={Overview of functional magnetic resonance imaging},
  author={Glover, Gary H},
  journal={Neurosurg. Clin.},
  volume={22},
  number={2},
  pages={133--139},
  year={2011},
  publisher={Elsevier}
}

@article{yu2018wide,
  title={Wide activation for efficient and accurate image super-resolution},
  author={Yu, Jiahui and Fan, Yuchen and Yang, Jianchao and Xu, Ning and Wang, Zhaowen and Wang, Xinchao and Huang, Thomas},
  journal={arXiv preprint arXiv:1808.08718},
  year={2018}
}

@article{wiaux2009compressed,
  title={Compressed sensing imaging techniques for radio interferometry},
  author={Wiaux, Yves and Jacques, Laurent and Puy, Gilles and Scaife, Anna MM and Vandergheynst, Pierre},
  journal={Mon. Not. R. Astron. Soc.},
  volume={395},
  number={3},
  pages={1733--1742},
  year={2009},
  publisher={The Royal Astronomical Society}
}

@inproceedings{chung2024decomposed,
  title={Decomposed Diffusion Sampler for Accelerating Large-Scale Inverse Problems},
  author={Chung, Hyungjin and Lee, Suhyeon and Ye, Jong Chul},
  booktitle={Proc. ICLR},
  year={May 2024}
}

@article{uecker2014espirit,
  title={{ESPIRiT}—an eigenvalue approach to autocalibrating parallel {MRI}: where {SENSE} meets {GRAPPA}},
  author={Uecker, Martin and Lai, Peng and Murphy, Mark J and Virtue, Patrick and Elad, Michael and Pauly, John M and Vasanawala, Shreyas S and Lustig, Michael},
  journal={Magn. Reson. Med.},
  volume={71},
  number={3},
  pages={990--1001},
  year={2014},
  publisher={Wiley Online Library}
}

@inproceedings{muckley20,
  author = {M. J. Muckley and R. Stern and T. Murrell and F. Knoll},
  title = {{TorchKbNufft}: A High-Level, Hardware-Agnostic Non-Uniform Fast {Fourier} Transform},
  booktitle = {ISMRM Workshop},
  year = {Jan. 2020},
}

@article{fessler2003nonuniform,
  title={{Nonuniform fast Fourier transforms} using min-max interpolation},
  author={Fessler, Jeffrey A and Sutton, Bradley P},
  journal={IEEE Trans. Signal Process.},
  volume={51},
  number={2},
  pages={560--574},
  year={2003},
  publisher={IEEE}
}

@inproceedings{aghabiglou2023deep,
  title={Deep network series for large-scale high-dynamic range imaging},
  author={Aghabiglou, Amir and Terris, Matthieu and Jackson, Adrian and Wiaux, Yves},
  booktitle={Proc. IEEE ICASSP},
  pages={1--5},
  year={Jun. 2023},
  organization={IEEE}
}

@article{aghabiglou2024r2d2,
  title={The {R2D2} deep neural network series paradigm for fast precision imaging in radio astronomy},
  author={Aghabiglou, Amir and San Chu, Chung and Dabbech, Arwa and Wiaux, Yves},
  journal={Astrophys. J. Suppl. Ser.},
  volume={273},
  number={1},
  pages={3},
  year={2024},
  publisher={IOP Publishing}
}

@article{aghabiglou2025r2d2,
  title={Toward a Robust {R2D2} Paradigm for Radio-interferometric Imaging: Revisiting Deep Neural Network Training and Architecture},
  author={Aghabiglou, Amir and Chu, Chung San and Tang, Chao and Dabbech, Arwa and Wiaux, Yves},
  journal={The Astrophysical Journal Supplement Series},
  volume={280},
  number={2},
  pages={63},
  year={2025},
  publisher={The American Astronomical Society}
}

@inproceedings{ronneberger2015u,
  title={U-net: Convolutional networks for biomedical image segmentation},
  author={Ronneberger, Olaf and Fischer, Philipp and Brox, Thomas},
  booktitle={Proc. MICCAI},
  pages={Part III 18, 234--241},
  organization={Springer},
  year={Oct. 2015}
}

@article{ramzi2022nc,
  title={{NC-PDNet}: A density-compensated unrolled network for {2D} and {3D non-Cartesian MRI} reconstruction},
  author={Ramzi, Zaccharie and Chaithya, GR and Starck, Jean-Luc and Ciuciu, Philippe},
  journal={IEEE Trans. Med. Imaging},
  volume={41},
  number={7},
  pages={1625--1638},
  year={2022},
  publisher={IEEE}
}

@article{qu2012undersampled,
  title={Undersampled {MRI} reconstruction with patch-based directional wavelets},
  author={Qu, Xiaobo and Guo, Di and Ning, Bende and Hou, Yingkun and Lin, Yulan and Cai, Shuhui and Chen, Zhong},
  journal={Magn. Reson. Imaging},
  volume={30},
  number={7},
  pages={964--977},
  year={2012},
  publisher={Elsevier}
}

@article{ahmad2020plug,
  title={Plug-and-play methods for magnetic resonance imaging: Using denoisers for image recovery},
  author={Ahmad, Rizwan and Bouman, Charles A and Buzzard, Gregery T and Chan, Stanley and Liu, Sizhuo and Reehorst, Edward T and Schniter, Philip},
  journal={IEEE Signal Process. Mag.},
  volume={37},
  number={1},
  pages={105--116},
  year={2020},
  publisher={IEEE}
}

@article{loshchilov2017fixing,
  title={Fixing weight decay regularization in adam},
  author={Loshchilov, Ilya and Hutter, Frank and others},
  journal={arXiv preprint arXiv:1711.05101},
  volume={5},
  year={2017}
}

@article{eo2018kiki,
  title={{KIKI-net}: cross-domain convolutional neural networks for reconstructing undersampled magnetic resonance images},
  author={Eo, Taejoon and Jun, Yohan and Kim, Taeseong and Jang, Jinseong and Lee, Ho-Joon and Hwang, Dosik},
  journal={Magn. Reson. Med.},
  volume={80},
  number={5},
  pages={2188--2201},
  year={2018},
  publisher={Wiley Online Library}
}

@article{han2019k,
  title={k-space deep learning for accelerated {MRI}},
  author={Han, Yoseo and Sunwoo, Leonard and Ye, Jong Chul},
  journal={IEEE Trans. Med. Imaging},
  volume={39},
  number={2},
  pages={377--386},
  year={2019},
  publisher={IEEE}
}

@article{stankovic20144d,
  title={{4D} flow imaging with {MRI}},
  author={Stankovic, Zoran and Allen, Bradley D and Garcia, Julio and Jarvis, Kelly B and Markl, Michael},
  journal={Cardiovasc. Diagn. Ther.},
  volume={4},
  number={2},
  pages={173},
  year="{2014}",
  publisher={AME Publications}
}

@article{adler2018learned,
  title={Learned primal-dual reconstruction},
  author={Adler, Jonas and {\"O}ktem, Ozan},
  journal={IEEE Trans. Med. Imaging},
  volume={37},
  number={6},
  pages={1322--1332},
  year={2018},
  publisher={IEEE}
}

@inproceedings{sriram2020end,
  title={End-to-end variational networks for accelerated {MRI} reconstruction},
  author={Sriram, Anuroop and Zbontar, Jure and Murrell, Tullie and Defazio, Aaron and Zitnick, C Lawrence and Yakubova, Nafissa and Knoll, Florian and Johnson, Patricia},
  booktitle={Proc. MICCAI},
  pages={Part II 23, 64--73},
  organization={Springer},
  year={Oct. 2020}
}

@article{pipe1999sampling,
  title={Sampling density compensation in {MRI}: rationale and an iterative numerical solution},
  author={Pipe, James G and Menon, Padmanabhan},
  journal={Magn. Reson. Med.},
  volume={41},
  number={1},
  pages={179--186},
  year={1999},
  publisher={Wiley Online Library}
}

@book{liang2000principles,
  title={Principles of magnetic resonance imaging},
  author={Liang, Zhi-Pei and Lauterbur, Paul C},
  year={2000},
  publisher={SPIE Optical Engineering Press Bellingham}
}

@article{lustig2007sparse,
  title={Sparse {MRI}: The application of compressed sensing for rapid {MR} imaging},
  author={Lustig, Michael and Donoho, David and Pauly, John M},
  journal={Magn. Reson. Med.},
  volume={58},
  number={6},
  pages={1182--1195},
  year={2007},
  publisher={Wiley Online Library}
}

@article{wilber2023scalable,
  title={Scalable precision wide-field imaging in radio interferometry: I. {uSARA} validated on {ASKAP} data},
  author={Wilber, Amanda G and Dabbech, Arwa and Jackson, Adrian and Wiaux, Yves},
  journal={Mon. Not. R. Astron. Soc.},
  volume={522},
  number={4},
  pages={5558--5575},
  year={2023},
  publisher={Oxford University Press}
}

@article{zbontar2018fastmri,
  title={{fastMRI}: An open dataset and benchmarks for accelerated {MRI}},
  author={Zbontar, Jure and Knoll, Florian and Sriram, Anuroop and Murrell, Tullie and Huang, Zhengnan and Muckley, Matthew J and Defazio, Aaron and Stern, Ruben and Johnson, Patricia and Bruno, Mary and others},
  journal={arXiv preprint arXiv:1811.08839},
  year={2018}
}

@article{zhang2023practical,
   author = {Zhang, Kai and Li, Yawei and Liang, Jingyun and Cao, Jiezhang and Zhang, Yulun and Tang, Hao and Fan, Deng-Ping and Timofte, Radu and Gool, Luc Van},
   title = {Practical Blind Image Denoising via {Swin-Conv-UNet} and Data Synthesis},
   journal = {Mach. Intell. Res.},
   DOI = {10.1007/s11633-023-1466-0},
   volume={20},
   number={6},
   pages={822--836},
   year={2023},
   publisher={Springer}
}

@article{kingma2014adam,
  title={Adam: A method for stochastic optimization},
  author={Kingma, Diederik P and Ba, Jimmy},
  journal={arXiv preprint arXiv:1412.6980},
  year={2014}
}

@article{mallat1993matching,
  title={Matching pursuits with time-frequency dictionaries},
  author={Mallat, St{\'e}phane G and Zhang, Zhifeng},
  journal={IEEE Trans. Signal Process.},
  volume={41},
  number={12},
  pages={3397--3415},
  year={1993},
  publisher={IEEE}
}

@article{paszke2019pytorch,
  title={Pytorch: An imperative style, high-performance deep learning library},
  author={Paszke, Adam and Gross, Sam and Massa, Francisco and Lerer and others},
  journal={Proc. NIPS},
  pages = {8024-8035},
  year={Dec. 2019}
}

@article{hyun2018deep,
  title={Deep learning for undersampled {MRI} reconstruction},
  author={Hyun, Chang Min and Kim, Hwa Pyung and Lee, Sung Min and Lee, Sungchul and Seo, Jin Keun},
  journal={Phys. Med. Biol.},
  volume={63},
  number={13},
  pages={135007},
  year={2018},
  publisher={IOP Publishing}
}

@article{guerquin2011fast,
  title={A fast wavelet-based reconstruction method for magnetic resonance imaging},
  author={Guerquin-Kern, Matthieu and Haberlin, M and Pruessmann, Klaas Paul and Unser, Michael},
  journal={IEEE Trans. Med. Imaging},
  volume={30},
  number={9},
  pages={1649--1660},
  year={2011},
  publisher={IEEE}
}

@article{block2007undersampled,
  title={Undersampled radial {MRI} with multiple coils. Iterative image reconstruction using a total variation constraint},
  author={Block, Kai Tobias and Uecker, Martin and Frahm, Jens},
  journal={Magn. Reson. Med.},
  volume={57},
  number={6},
  pages={1086--1098},
  year={2007},
  publisher={Wiley Online Library}
}

@article{pruessmann1999sense,
  title={{SENSE}: sensitivity encoding for fast {MRI}},
  author={Pruessmann, Klaas P and Weiger, Markus and Scheidegger, Markus B and Boesiger, Peter},
  journal={Magn. Reson. Med.},
  volume={42},
  number={5},
  pages={952--962},
  year={1999},
  publisher={Wiley Online Library}
}

@article{dhariwal2021diffusion,
  title={Diffusion models beat {GANs} on image synthesis},
  author={Dhariwal, Prafulla and Nichol, Alexander},
  journal={Proc. NIPS},
  volume={34},
  pages={8780--8794},
  year={Dec. 2021}
}

@inproceedings{song2020denoising,
  title={Denoising diffusion implicit models},
  author={Song, Jiaming and Meng, Chenlin and Ermon, Stefano},
  booktitle={Proc. ICLR},
  year={May 2021}
}

@article{knoll2020deep,
  title={Deep-learning methods for parallel magnetic resonance imaging reconstruction: A survey of the current approaches, trends, and issues},
  author={Knoll, Florian and Hammernik, Kerstin and Zhang, Chi and Moeller, Steen and Pock, Thomas and Sodickson, Daniel K and Akcakaya, Mehmet},
  journal={IEEE Signal Process. Mag.},
  volume={37},
  number={1},
  pages={128--140},
  year={2020},
  publisher={IEEE}
}

@inproceedings{zhu2023denoising,
  title={Denoising diffusion models for plug-and-play image restoration},
  author={Zhu, Yuanzhi and Zhang, Kai and Liang, Jingyun and Cao, Jiezhang and Wen, Bihan and Timofte, Radu and Van Gool, Luc},
  booktitle={Proceedings of the IEEE/CVF conference on computer vision and pattern recognition},
  pages={1219--1229},
  year={2023}
}

@misc{sigpy,
  author       = {Ong, Frank and Lustig, Michael},
  title        = {{SigPy}: A Python Package for High Performance Iterative Reconstruction},
  year         = {2024},
  howpublished = {\url{https://github.com/mikgroup/sigpy}},
  note         = {Version 0.1.27}
}

@article{daval2023deep,
  title={Deep learning-assisted model-based off-resonance correction for {non-Cartesian SWI}},
  author={Daval-Fr{\'e}rot, Guillaume and Massire, Aur{\'e}lien and Mailh{\'e}, Boris and Nadar, Mariappan and Bapst, Blanche and Luciani, Alain and Vignaud, Alexandre and Ciuciu, Philippe},
  journal={Magnetic Resonance in Medicine},
  volume={90},
  number={4},
  pages={1431--1445},
  year={2023},
  publisher={Wiley Online Library}
}

@INPROCEEDINGS{NCPDNET2025,
  author={Tanabene, A. and G. R., Chaithya and Massire, A. and Nadar, M. and Ciuciu, P.},
  booktitle={2025 IEEE 22nd International Symposium on Biomedical Imaging (ISBI)}, 
  title={Benchmarking 3D Multi-Coil {NC-PDNet MRI} Reconstruction}, 
  year={2025},
  volume={},
  number={},
  pages={1-5},
  doi={10.1109/ISBI60581.2025.10980654}}

@article{ying2007joint,
  title={Joint image reconstruction and sensitivity estimation in {SENSE (JSENSE)}},
  author={Ying, Leslie and Sheng, Jinhua},
  journal={Magnetic Resonance in Medicine: An Official Journal of the International Society for Magnetic Resonance in Medicine},
  volume={57},
  number={6},
  pages={1196--1202},
  year={2007},
  publisher={Wiley Online Library}
}

@inproceedings{he2015delving,
  title={Delving deep into rectifiers: Surpassing human-level performance on imagenet classification},
  author={He, Kaiming and Zhang, Xiangyu and Ren, Shaoqing and Sun, Jian},
  booktitle={Proceedings of the IEEE international conference on computer vision},
  pages={1026--1034},
  year={2015}
}

@article{repetti2017non,
  title={Non-convex optimization for self-calibration of direction-dependent effects in radio interferometric imaging},
  author={Repetti, Audrey and Birdi, Jasleen and Dabbech, Arwa and Wiaux, Yves},
  journal={Monthly Notices of the Royal Astronomical Society},
  volume={470},
  number={4},
  pages={3981--4006},
  year={2017},
  publisher={Oxford University Press}
}

@article{chouzenoux2016block,
  title={A block coordinate variable metric forward--backward algorithm},
  author={Chouzenoux, Emilie and Pesquet, Jean-Christophe and Repetti, Audrey},
  journal={Journal of Global Optimization},
  volume={66},
  number={3},
  pages={457--485},
  year={2016},
  publisher={Springer}
}

@incollection{bauschke2020correction,
  title={Correction to: convex analysis and monotone operator theory in Hilbert spaces},
  author={Bauschke, Heinz H and Combettes, Patrick L},
  booktitle={Convex analysis and monotone operator theory in Hilbert spaces},
  pages={C1--C4},
  year={2020},
  publisher={Springer}
}

\end{document}